\pdfoutput=1
\documentclass[aps,pra,reprint,superscriptaddress,amsmath,amssymb,floatfix,longbibliography]{revtex4-2}

\usepackage{bm}
\usepackage{graphicx}
\usepackage{mathtools}
\usepackage{physics}
\usepackage{microtype}
\usepackage[colorlinks=true,linkcolor=blue,citecolor=blue,urlcolor=blue]{hyperref}

\begin{document}

\title{Quantum relative entropy and spatial information loss in coherently seeded, spacelike-separated biphoton sources}

\author{Tai Hyun Yoon}
\email{thyoon@korea.ac.kr}
\affiliation{Department of Physics, Korea University, Anamro 145, Seongbukgu, Seoul 02841, Republic of Korea}
\affiliation{Center for Molecular Spectroscopy and Dynamics, Institute for Basic Science (IBS), Anamro 145, Seongbukgu, Seoul 02841, Republic of Korea}

\date{\today}

\begin{abstract}
We develop an open-system description of two spacelike-separated, coherently seeded parametric down-conversion sources in which quantum relative entropy plays the organizing role that it has in recent algebraic quantum field theory, where the relative entropy between the vacuum and a coherent excitation governs the energy flux across a causal horizon [P.~Dorau and A.~Much, Phys.~Rev.~Lett.~\textbf{136}, 091602 (2026)]. Two periodically poled lithium niobate crystals are pumped by an optical frequency comb and seeded by a continuous-wave laser in a coherent state $|\alpha\rangle$, so that each signal photon is entangled with an idler single-photon-added coherent state. Tracing over the unmeasured idler modes gives a spatial Lindblad master equation along the propagation axis whose dephasing rate, $\Gamma_{\mathrm{deph}} = -(1/L_c)\ln F$, is fixed by the overlap $F = |\alpha|^2/(1+|\alpha|^2)$ of the two idler states over the interaction length $L_c$. The decay of the relative entropy of the signal state with respect to its dephased mixture defines a non-negative information loss rate per unit length, whose integral over $L_c$ is the binary entropy of $(1+F)/2$. The same overlap saturates the wave-particle-entanglement triality $P^2+E^2+V^2=1$. Combined with the quotient-space construction of the tensor product, the model describes the observed single-photon interference and its three equivalent phase scans as a local projection onto a detector-defined coset, with no action on the remote idler fields. Finally, the symplectic form that underlies the relative entropy is that of the time-frequency Weyl displacements generating Gottesman-Kitaev-Preskill stabilizers on the comb, and the stabilizer readout contrast is limited by $F$.
\end{abstract}

\maketitle

%=====================================================================
\section{Introduction}
\label{sec:intro}
%=====================================================================

How definite measurement outcomes and first-order coherence emerge from composite entangled states prepared across spacelike-separated regions, without an \textit{ad hoc} collapse postulate or superluminal signaling, remains a central question in the foundations of quantum mechanics~\cite{Bohr1928,Einstein1935,Englert1996,Bell1964,Brunner2014,Wiseman2010}. In parallel, work in relativistic quantum field theory (QFT) has shown that quantum-information quantities govern field and spacetime dynamics~\cite{Jacobson1995,Casini2008,Hollands2019,Kurpicz2021}. Most recently, Dorau and Much~\cite{Dorau2026} showed within algebraic QFT (AQFT) that the semiclassical Einstein equations follow from the Araki-Uhlmann quantum relative entropy~\cite{Araki1976,Uhlmann1977} evaluated between a vacuum reference state $\omega_0$ and a coherent field excitation $\omega_\phi$ across a local causal horizon.

The reason relative entropy, and not entanglement entropy, appears in that result is structural. In continuous-mode QFT the von Neumann entanglement entropy of a region is ultraviolet divergent unless a mode cutoff is imposed. The relative entropy $S^{\mathrm{rel}}(\omega_0 \parallel \omega_\phi)$, in contrast, is finite, is given by a symplectic Noether charge flux, and quantifies the \emph{entanglement deficit} created when a coherent excitation displaces the entangled vacuum~\cite{Dorau2026,Casini2019,Longo2019}. A dictionary between the AQFT quantities and those of quantum optics is given in Appendix~\ref{app:AQFT_QED_bridge}.

\begin{figure*}[t]
\centering
\includegraphics[width=0.9\textwidth]{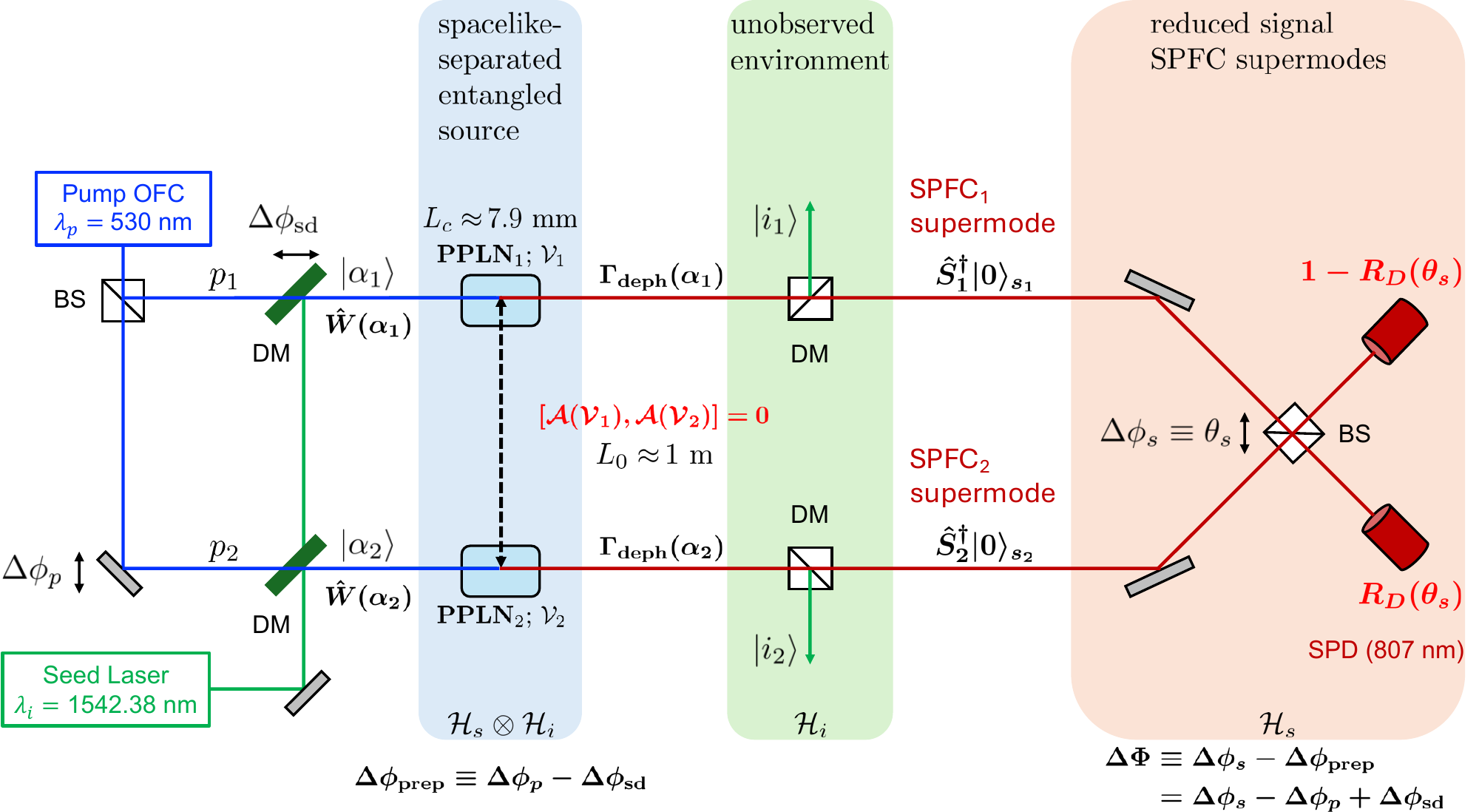}
\caption{Simplified schematic of the dual-crystal entangled nonlinear biphoton source (ENBS)~\cite{Lee2018,Yoon2021,Yoon2026c,Yoon2026d}. The full optical layout, including the corner-cube delay lines that equalize the two pump paths and the two signal paths, is given in Fig.~1(a) of Ref.~\cite{Lee2018}. Two periodically poled lithium niobate crystals ($\mathrm{PPLN}_1$ and $\mathrm{PPLN}_2$) with interaction length $L_c \approx 7.9~\mathrm{mm}$ are separated by a baseline $L_0 \approx 1~\mathrm{m}$~\cite{Lee2018,Yoon2021}, so that the local algebras of the spacetime interaction volumes $\mathcal{V}_1$ and $\mathcal{V}_2$ commute, $[\mathcal{A}(\mathcal{V}_1), \mathcal{A}(\mathcal{V}_2)] = 0$~\cite{Yoon2026d}. The pump optical frequency comb ($\lambda_p = 530~\mathrm{nm}$) is divided at the input beam splitter (BS) into the arms $p_1$ and $p_2$, and the continuous-wave seed laser ($\lambda_i = 1542.38~\mathrm{nm}$) is combined with each pump arm on a dichroic mirror (DM), so that the crystals are seeded in coherent states $|\alpha_1\rangle$ and $|\alpha_2\rangle$ prepared by independent Weyl displacements $\hat{W}(\alpha_1)$ and $\hat{W}(\alpha_2)$~\cite{Lee2018,Yoon2021,Yoon2026c}; this fixes the preparation phase $\Delta\phi_{\mathrm{prep}} \equiv \Delta\phi_p - \Delta\phi_{\mathrm{sd}}$. The crystals emit single-photon frequency-comb (SPFC) signal supermodes $\hat{S}_1^\dagger|0\rangle_{s_1}$ and $\hat{S}_2^\dagger|0\rangle_{s_2}$ entangled with idler single-photon-added coherent states (SPACS) in $\mathcal{H}_s \otimes \mathcal{H}_i$~\cite{Lee2018,Yoon2021,Yoon2026c}. Behind each crystal a DM removes the pump and idler fields from the signal path. Tracing out the unobserved idler modes ($|i_1\rangle$, $|i_2\rangle$) gives the local spatial Lindblad dephasing rates $\Gamma_{\mathrm{deph}}(\alpha_1)$ and $\Gamma_{\mathrm{deph}}(\alpha_2)$, defined below Eq.~\eqref{eq:gamma_deph}, that depend only on the local seed of each crystal~\cite{Yoon2021,Yoon2026d}. Only the two signal supermodes are combined at the output BS; with readout phase $\Delta\phi_s \equiv \theta_s$ this projects the composite state onto detector-defined coset slices. The counting rates at the two output ports, $R_D(\theta_s)$ and $1-R_D(\theta_s)$, are normalized to unit total rate, with $R_D \equiv R_B$ the bright-port rate of Eq.~\eqref{eq:count_rate}. The rates at the 807-nm single-photon-sensitive detectors (an EMCCD behind a grating monochromator in Ref.~\cite{Lee2018}) depend only on the total phase $\Delta\Phi \equiv \Delta\phi_s - \Delta\phi_{\mathrm{prep}} = \Delta\phi_s - \Delta\phi_p + \Delta\phi_{\mathrm{sd}}$~\cite{Yoon2026b,Yoon2026c}.}
\label{fig:fig1}
\end{figure*}

This raises the question addressed here: is there a laboratory quantum electrodynamics (QED) system in which the same relative-entropy structure can be written down explicitly and compared with measured quantities? Nonlinear interferometry based on induced coherence without induced emission~\cite{Zou1991,Wang1991,Ou1990,Heuer2015} and coherently seeded entangled nonlinear biphoton sources (ENBS)~\cite{Lee2018,Yoon2021} provide such a system. As illustrated in Fig.~\ref{fig:fig1}, two spacelike-separated periodically poled lithium niobate (PPLN) crystals are pumped by the two arms of the same phase-locked optical frequency comb (OFC)~\cite{Hansch2006,Hall2006} and seeded by the same narrow-linewidth continuous-wave (cw) laser. Each crystal then emits a broadband single-photon frequency-comb (SPFC) supermode that is entangled with a single-photon-added coherent state (SPACS) in the idler mode~\cite{Lee2018,Yoon2021,Zavatta2004}. Earlier theoretical work on this dual-crystal architecture has treated quantum-vacuum control of time qubits~\cite{Yoon2026c}, the measurement-defined equivalence of three phase scans~\cite{Yoon2026b}, nonlocal state reduction in terms of free-vector-space quotient projections and open-system Lindblad dynamics~\cite{Scholes2026c,Yoon2026d}, and hardware-native time-frequency Gottesman-Kitaev-Preskill (TF-GKP) logical qubits~\cite{Gottesman2001,Fluhmann2019,Fabre2020,Descamps2024,Yoon2026a}.

In this paper we formulate the dual-crystal ENBS in the language of local Weyl algebras and open quantum systems, with quantum relative entropy and its rate of loss along the propagation axis as the organizing quantities. The results are as follows. (i) Injecting a cw coherent idler seed $|\alpha\rangle$ acts as a Weyl displacement of the multimode QED vacuum. The displacement creates an entanglement deficit that suppresses which-path information and restores the single-photon interference visibility $V = |\alpha|^2/(1+|\alpha|^2)$~\cite{Yoon2021,Yoon2026b,Yoon2026c}. (ii) The reduced signal state obeys a spatial Lindblad dephasing master equation along the crystal axis~\cite{Yoon2026d}, and the decay of its relative entropy with respect to the fully dephased mixture defines a non-negative information loss rate that we evaluate in closed form. (iii) The local detection of the signal photon selects a coset representative of the composite state and leaves the remote idler fields untouched, which accounts for the equivalence of the pump, seed, and readout phase scans. (iv) The symplectic form underlying the relative entropy defines the commuting displacement stabilizers of TF-GKP states on the comb~\cite{Yoon2026a,Yoon2026c}.

The paper is organized as follows. Section~\ref{sec:algebras} introduces the local field algebras, the comb supermodes, and the seeded joint state. Section~\ref{sec:relent} relates the SPACS overlap to the relative entropy of the coherent seed and to the triality relation. Section~\ref{sec:lindblad} derives the information loss rate associated with the spatial Lindblad equation. Section~\ref{sec:quotient} treats the measurement as a quotient-space projection. Section~\ref{sec:comparison} compares the model with the measurements reported in Ref.~\cite{Lee2018}. Section~\ref{sec:gkp} connects the symplectic structure to TF-GKP stabilizers, and Sec.~\ref{sec:conclusion} discusses the scope and limits of the results.

%=====================================================================
\section{Local QED algebras and coherent Weyl excitations}
\label{sec:algebras}
%=====================================================================

\subsection{Geometry and microcausality}
\label{subsec:geometry}

Consider two identical type-0 PPLN crystals ($\mathrm{PPLN}_1$ and $\mathrm{PPLN}_2$ in Fig.~\ref{fig:fig1}) with a poling period of $7.3~\mu\mathrm{m}$, operated at a phase-matching temperature of $121.5\,^{\circ}\mathrm{C}$~\cite{Lee2018}. Each crystal has several poled channels, of which a single channel of length $15~\mathrm{mm}$ is used. The pump and seed beams are focused into this channel with lenses of focal length $100~\mathrm{mm}$ to a waist $w_0 \approx 17.3~\mu\mathrm{m}$. The resulting confocal parameter, $b = 2z_R \approx 7.9~\mathrm{mm}$ with $z_R$ the Rayleigh range inside the crystal, sets the effective interaction length $L_c \approx 7.9~\mathrm{mm}$. The two crystals are separated by a baseline $L_0 \approx 1~\mathrm{m}$ in the laboratory frame. The 530-nm pump comb is the second harmonic of a 1060-nm fundamental comb with a spectral bandwidth of 46~nm, whose repetition rate $f_r = \Omega_{\mathrm{rep}}/2\pi = 250.0~\mathrm{MHz}$ and carrier-envelope offset frequency of 20.0~MHz are phase-locked to a GPS-disciplined Rb atomic clock. The second harmonic is generated in a noncritically phase-matched LBO crystal, whose acceptance bandwidth limits the pump spectrum to about 3~nm. The pump pulse width, estimated from the width of the fringe-visibility envelope as a function of the pump path difference, is $\tau_p \approx 8.4~\mathrm{ps}$ (FWHM)~\cite{Lee2018}. The cw seed at $1542.384~\mathrm{nm}$ has a linewidth of 1~Hz.

Two further time scales must be distinguished from $\tau_p$. The first is the interaction window $\Delta\tau_{\mathrm{int}}$, the laboratory time during which the pump pulse is inside the crystal. For the full 15-mm channel it is at most $\Delta\tau_{\mathrm{int}} \approx n_g L/c + \tau_p \approx 130~\mathrm{ps}$, where $L = 15~\mathrm{mm}$ and $n_g \approx 2.4$ is the pump group index; most pairs are created within the shorter length $L_c$ around the focus. A pair can be created in a given crystal at any time within this window, and it is this window that enters the causal argument below. It is much shorter than $L_0/c \approx 3.3~\mathrm{ns}$. The second is the duration $\tau_s$ of the emitted signal wave packet, which is set by the bandwidth of the seeded signal comb. Spontaneous emission has a bandwidth of more than 20~nm at 807~nm, whereas with the cw idler seed the signal comb has an optical bandwidth $\Delta\nu_s \approx 52.3~\mathrm{GHz}$, that is, about 209 comb teeth~\cite{Lee2018}. For a transform-limited Gaussian pulse with $\tau_s\,\Delta\nu_s = 0.441$ this corresponds to a signal duration $\tau_s \approx 8.4~\mathrm{ps}$ and a length $c\tau_s \approx 2.5~\mathrm{mm}$.

The optical layout is a hybrid interferometer with a classical input side and a quantum output side. Figure~\ref{fig:fig1} is a simplified schematic; the full layout, including the path-equalizing delay lines, is shown in Fig.~1(a) of Ref.~\cite{Lee2018}. On the input side, the pump OFC is divided at a beam splitter into two arms $p_1$ and $p_2$, and the cw seed is divided at a second beam splitter. In each arm the seed is combined with the pump on a dichroic mirror (DM), so that the pump and seed fields enter each crystal collinearly in the same spatial mode, and the signal generated in each crystal leaves in that mode. A corner-cube retroreflector on a translation stage in one pump arm equalizes the two pump path lengths, the fringe visibility being maximal when the path-length differences of both the pump and the signal beams are zero~\cite{Lee2018}, and sets the pump phase difference $\Delta\phi_p$, and a translatable mirror in one seed arm sets the seed phase difference $\Delta\phi_{\mathrm{sd}}$. Behind each crystal a second DM removes the pump and the seed (idler) fields from the signal path. The idler modes $|i_1\rangle$ and $|i_2\rangle$ are neither recombined nor detected, and they constitute the unobserved environment. On the output side, only the two signal SPFC supermodes are combined, at the output beam splitter (BS). The first-order interference of the signal supermodes is recorded with an electron-multiplying CCD (EMCCD), a single-photon-sensitive array detector, placed behind a grating monochromator, which gives the spectrally resolved photon-counting rate at 807~nm~\cite{Lee2018}. Each data point is integrated over 10~ms, that is, over $2.5\times 10^{6}$ pump pulses. About $5\times 10^{4}$ signal photons are detected during one exposure, which corresponds to about one photon per 50 pump pulses from each crystal, a filling factor of 0.02 per pulse~\cite{Lee2018}; the fringes of Fig.~2(a) of Ref.~\cite{Lee2018} have maxima of about $2.6\times 10^{4}$ photons per 10~ms. A mean photon number of 0.02 per pulse justifies the single-photon description used below. The phases $\Delta\phi_p$, $\Delta\phi_{\mathrm{sd}}$, and $\Delta\phi_s$ are scanned slowly with piezo-driven translation stages that carry the corner-cube retroreflectors or mirrors, and each phase is treated as stationary during one 10-ms integration. No pump or seed light closes an interferometer at the output BS, so the three phases $\Delta\phi_p$, $\Delta\phi_{\mathrm{sd}}$, and $\Delta\phi_s$ are set by three different optical elements.

The signal amplitudes from the two crystals must overlap in time at the output BS, to within the length of the signal supermode, $c\tau_s \approx 2.5~\mathrm{mm}$. A second corner-cube retroreflector on a translation stage in one signal arm equalizes the two signal path lengths and sets the readout phase $\Delta\phi_s \equiv \theta_s$~\cite{Lee2018}.

Let $\ell_{p,j}$ denote the pump path length from the input beam splitter to $\mathrm{PPLN}_j$, so that the pump arrival times at the two crystals differ by $(\ell_{p,2} - \ell_{p,1})/c$. Pair-creation events driven by the same pump pulse are spacelike separated provided
\begin{equation}
|\ell_{p,1} - \ell_{p,2}| + c\,\Delta\tau_{\mathrm{int}} < L_0 ,
\label{eq:spacelike_condition}
\end{equation}
up to corrections of order $L_c$, where $L_0$ is the straight-line distance between the crystals and $c\,\Delta\tau_{\mathrm{int}} \approx 4~\mathrm{cm}$. With the pump path lengths equalized, $|\ell_{p,1} - \ell_{p,2}| \ll c\,\Delta\tau_{\mathrm{int}}$, the two crystals are pumped simultaneously in the laboratory frame and Eq.~\eqref{eq:spacelike_condition} holds with a wide margin. The spacetime interaction volumes $\mathcal{V}_1$ and $\mathcal{V}_2$ then satisfy $(x_1 - x_2)^2 \equiv |\mathbf{r}_1 - \mathbf{r}_2|^2 - c^2 (t_1 - t_2)^2 > 0$ for all events $x_1 = (t_1, \mathbf{r}_1) \in \mathcal{V}_1$ and $x_2 = (t_2, \mathbf{r}_2) \in \mathcal{V}_2$. Quantizing the transverse electromagnetic vector potential $\hat{\mathbf{A}}_\perp(x)$ in the Coulomb gauge, the gauge-invariant field observables satisfy microcausality~\cite{Peskin1995,Yoon2026d},
\begin{equation}
\big[\mathcal{A}(\mathcal{V}_1), \mathcal{A}(\mathcal{V}_2)\big] = 0,
\label{eq:microcausality}
\end{equation}
where $\mathcal{A}(\mathcal{V}_j)$ denotes the local algebra of QED field observables localized in $\mathrm{PPLN}_j$ ($j \in \{1,2\}$).

\subsection{Comb supermodes and the Weyl algebra}
\label{subsec:weyl}

Each crystal is driven by a pump OFC with teeth at $\omega_{p,n} = \omega_{p,0} + n\Omega_{\mathrm{rep}}$~\cite{Lee2018,Hansch2006,Hall2006} and by a shared monochromatic cw idler seed at frequency $\omega_{\mathrm{sd}}$ with coherent amplitudes $\alpha_j = |\alpha_j|e^{i\phi_{\mathrm{sd},j}}$. Each crystal then generates a collective SPFC signal supermode~\cite{Lee2018,Yoon2026a,Yoon2026c},
\begin{equation}
\hat{S}_j^\dagger = \sum_{n=0}^{N-1} f_n \hat{a}_{s,j}^\dagger(\omega_{s,n}), \quad \sum_{n=0}^{N-1} |f_n|^2 = 1,
\label{eq:SPFC_supermode}
\end{equation}
with $[\hat{S}_j, \hat{S}_k^\dagger] = \delta_{jk}$. Because the seed fixes the idler frequency, each signal tooth is tied to one pump tooth, $\omega_{s,n} = \omega_{p,n} - \omega_{\mathrm{sd}}$, and the amplitudes $f_n$ are set by the pump spectrum and the phase-matching function of the crystal. For the seeded signal bandwidth $\Delta\nu_s \approx 52.3~\mathrm{GHz}$ of Sec.~\ref{subsec:geometry} and $f_r = 250~\mathrm{MHz}$, the supermode of Eq.~\eqref{eq:SPFC_supermode} comprises $N = \Delta\nu_s/f_r \approx 209$ teeth~\cite{Lee2018}.

In continuous-mode QED, as in AQFT~\cite{Dorau2026,Kurpicz2021,Kay1991}, the local bosonic field algebra is built in three steps. First, restricting the field to the single focused spatial mode in which the pump, seed, and signal propagate collinearly through the crystal, referred to below as the collinear mode, leaves a field that depends on a single coordinate, the retarded time $U = t - z/c$. We write $\hat{\Phi}(U)$ for the field operator at the point $U$. Second, the field at a sharp point is not a well-defined operator, and it is averaged over $U$ with a real weight function $\phi(U)$, the wave-packet profile, which is an ordinary function and not an operator,
\begin{equation}
\hat{\Phi}(\phi) = \int \hat{\Phi}(U)\,\phi(U)\,dU .
\label{eq:smeared}
\end{equation}
Here $U$ is the integration variable, so the smeared field $\hat{\Phi}(\phi)$ does not depend on $U$. It is a single Hermitian operator labeled by the function $\phi$; the same symbol $\hat{\Phi}$ with a function as its argument denotes the smeared field. Third, the unitary Weyl operator associated with the profile $\phi$ is
\begin{equation}
\hat{W}(\phi) = \exp\!\big[i\hat{\Phi}(\phi)\big].
\label{eq:Weyl_def}
\end{equation}

The algebra of these operators is fixed by the symplectic form, which for two profiles $\phi$ and $\psi$ reads
\begin{equation}
\sigma(\phi, \psi) = \int \big(\phi\,\partial_U\psi - \psi\,\partial_U\phi\big)\,dU .
\label{eq:symplectic_form}
\end{equation}
The smeared fields satisfy $[\hat{\Phi}(\phi), \hat{\Phi}(\psi)] = i\sigma(\phi, \psi)$, so that the Weyl operators obey the composition law~\cite{Dorau2026}
\begin{equation}
\hat{W}(\phi)\hat{W}(\psi) = e^{-\frac{i}{2}\sigma(\phi,\psi)}\,\hat{W}(\phi+\psi).
\label{eq:Weyl_product}
\end{equation}
Equations~\eqref{eq:smeared}--\eqref{eq:Weyl_product} are written in natural units, $\hbar = c = 1$, with the field and the profiles scaled so that $\hat{\Phi}(\phi)$ and $\sigma(\phi,\psi)$ are dimensionless. All relations between measurable quantities below are given in SI units.

\subsection{Seeded joint state}
\label{subsec:joint}

With the pump treated as a classical field, the $\chi^{(2)}$ parametric interaction in crystal $j$ is
\begin{equation}
\hat{H}_j = i\hbar\big(\kappa_j\, \hat{S}_j^\dagger \hat{a}_{i,j}^\dagger - \kappa_j^*\, \hat{S}_j \hat{a}_{i,j}\big),
\label{eq:H_chi2}
\end{equation}
where $\hat{a}_{i,j}$ is the idler-mode operator and the coupling rate $\kappa_j = |\kappa| e^{i\phi_{p,j}}$ is proportional to $\chi^{(2)}$ and to the pump amplitude, whose phase is $\phi_{p,j}$. Writing the idler operator as $\hat{a}_{i,j} = \alpha_j + \delta\hat{a}_{i,j}$ separates the classical seed amplitude $\alpha_j$ from the vacuum fluctuations $\delta\hat{a}_{i,j}$. Equation~\eqref{eq:H_chi2} then splits into a phase-insensitive spontaneous two-mode squeezing part $\hat{H}_0 = i\hbar(\kappa\, \hat{S}^\dagger \delta\hat{a}_i^\dagger - \kappa^* \hat{S}\, \delta\hat{a}_i)$ and a phase-sensitive stimulated displacement generator $\hat{V}$~\cite{Yoon2026c},
\begin{equation}
\hat{H}_\pm = \hat{H}_0 \pm \hat{V}, \quad \hat{V} = i\hbar\big(g_0 \hat{S}^\dagger - g_0^* \hat{S}\big), \quad g_0 = \kappa\,\alpha^* ,
\label{eq:H_branch}
\end{equation}
where the crystal index is suppressed. The stimulated coupling rate $g_0$ has magnitude $|\kappa||\alpha|$ and phase $\phi_{p,j} - \phi_{\mathrm{sd},j}$, which is the three-wave creation phase $\phi_{s,j}^{(0)} = \phi_{p,j} - \phi_{\mathrm{sd},j}$ of the signal. Setting their difference to $\Delta\phi_s^{(0)} = \pi$ flips $\hat{V} \to -\hat{V}$ between the arms $j=1,2$ and defines the two-dimensional path-qubit subspace $\mathcal{H}_T = \mathrm{span}\{|+\rangle, |-\rangle\}$, the time qubit of Refs.~\cite{Bloch2025,Yoon2026c}.

We take the creation-stage phase differences as arm 1 minus arm 2, $\Delta\phi_p \equiv \phi_{p,1} - \phi_{p,2}$ and $\Delta\phi_{\mathrm{sd}} \equiv \phi_{\mathrm{sd},1} - \phi_{\mathrm{sd},2}$, so that the signal amplitude of the second crystal carries the relative phase $e^{-i\Delta\phi_{\mathrm{prep}}}$ with $\Delta\phi_{\mathrm{prep}} \equiv \Delta\phi_p - \Delta\phi_{\mathrm{sd}}$. In the single-photon regime, the joint signal-idler state prepared across the two crystals is
\begin{equation}
|\Psi_{si}\rangle = c_1 |1,0\rangle_s |m_1\rangle_i + c_2 e^{-i\Delta\phi_{\mathrm{prep}}} |0,1\rangle_s |m_2\rangle_i,
\label{eq:joint_state}
\end{equation}
where $|1,0\rangle_s \equiv \hat{S}_1^\dagger|0\rangle_s \equiv |+\rangle$, $|0,1\rangle_s \equiv \hat{S}_2^\dagger|0\rangle_s \equiv |-\rangle$, and $|c_1|^2 + |c_2|^2 = 1$. The idler states $|m_j\rangle_i$ are normalized SPACS, obtained by acting with $\hat{a}_{i,j}^\dagger$ on the coherent states $|\alpha_j\rangle = \hat{W}(\alpha_j)|0\rangle_i$~\cite{Yoon2021,Zavatta2004}. Here $\hat{W}(\alpha_j)$ is the Weyl operator of Eqs.~\eqref{eq:smeared} and \eqref{eq:Weyl_def} for the seed wave packet $\phi_j(U)$ of crystal $j$,
\begin{align}
\hat{W}(\alpha_j) \equiv \hat{W}(\phi_j) &= \exp\!\left[i\int \hat{\Phi}(U)\,\phi_j(U)\,dU\right] \nonumber\\
&= \exp\!\big(\alpha_j \hat{a}_{i,j}^\dagger - \alpha_j^* \hat{a}_{i,j}\big).
\label{eq:W_alpha}
\end{align}
The second line follows from the decomposition $\hat{\Phi}(U) = \hat{\Phi}^{(+)}(U) + \hat{\Phi}^{(-)}(U)$ of the field into its annihilation and creation parts. For a wave packet confined to the seeded idler mode with normalized mode function $u_j(U)$, $\hat{\Phi}^{(+)}(U) \to u_j(U)\,\hat{a}_{i,j}$, so that $\int \hat{\Phi}^{(+)}(U)\,\phi_j(U)\,dU = i\alpha_j^*\,\hat{a}_{i,j}$ with the seed amplitude defined by
\begin{equation}
\alpha_j^* = -i\int u_j(U)\,\phi_j(U)\,dU .
\label{eq:alpha_def}
\end{equation}
The smeared field is then $\hat{\Phi}(\phi_j) = -i(\alpha_j \hat{a}_{i,j}^\dagger - \alpha_j^* \hat{a}_{i,j})$, and $\hat{W}(\alpha_j)$ is the displacement operator of the seeded idler mode. The normalized SPACS are
\begin{equation}
|m_j\rangle_i = \frac{e^{i\phi_{\mathrm{sd},j}}}{\sqrt{1+|\alpha_j|^2}}\,\hat{a}_{i,j}^\dagger |\alpha_j\rangle_{i,j}.
\label{eq:SPACS}
\end{equation}
The phase factor $e^{i\phi_{\mathrm{sd},j}}$ refers each SPACS to the phase of its own seed. With this convention the overlap $\langle m_2 | m_1 \rangle_i$ of the two idler states is real and positive, and the relative seed phase appears only in the preparation phase $\Delta\phi_{\mathrm{prep}}$ of Eq.~\eqref{eq:joint_state}, in agreement with the phase $\phi_{p,j} - \phi_{\mathrm{sd},j}$ of the stimulated coupling $g_0$.

Two limits of Eq.~\eqref{eq:joint_state} are familiar. In the symmetric, unseeded limit ($|c_1| = |c_2| = 1/\sqrt{2}$ and $\alpha_j = 0$) it is the maximally entangled bipartite state of the kind considered by Einstein, Podolsky, and Rosen~\cite{Einstein1935} and Bell~\cite{Bell1964} and tested in the experiments of Clauser, Aspect, and Zeilinger~\cite{Freedman1972,Aspect1982,Weihs1998}. Under strong symmetric seeding ($|\alpha_j| = |\alpha| \gg 1$) the idler states become indistinguishable, the two subsystems decouple, and the signal field approaches the pure superposition $(|1,0\rangle_s + e^{-i\Delta\phi_{\mathrm{prep}}} |0,1\rangle_s)/\sqrt{2}$~\cite{Zou1991,Yoon2021,Yoon2026c}.

%=====================================================================
\section{Quantum relative entropy and entanglement deficit}
\label{sec:relent}
%=====================================================================

\subsection{Relative entropy of a coherent excitation}
\label{subsec:srel_aqft}

The result of Dorau and Much~\cite{Dorau2026}, which builds on modular theory~\cite{Takesaki1970,Bisognano1975}, can be stated in quantum-optical terms. The reference state $\omega_0$ is the vacuum, and $\omega_\phi$ is the coherent state $\hat{W}(\phi)|0\rangle$ obtained by displacing the vacuum with the wave packet $\phi$ (Appendix~\ref{app:AQFT_QED_bridge}). The two states are compared on a null sheet $\mathcal{H}_B^{\mathcal{R}}$, the three-dimensional surface swept out by a light front. A point on this surface is labeled by the retarded time $U = t - z/c$, measured from the edge $U = 0$ of the region, and by its position on the two-dimensional cross section $\mathcal{S}$ transverse to the propagation direction. The area element of $\mathcal{S}$ is $d\mathrm{vol}_{\mathcal{S}}$, equal to $dx\,dy$ for propagation along $z$. The relative entropy is
\begin{align}
S^{\mathrm{rel}}(\omega_0 \parallel \omega_\phi) &= \frac{1}{2}\frac{d}{d\lambda}\Big|_{\lambda=0}\sigma(\phi^\lambda,\phi) \nonumber\\
&= 2\pi \int_{\mathcal{H}_B^{\mathcal{R}}} |U|\,\langle :T_{UU}: \rangle_{\omega_\phi}\,dU\,d\mathrm{vol}_{\mathcal{S}}.
\label{eq:Dorau_Srel}
\end{align}
In the first line, $\phi^\lambda(U) = \phi(e^{2\pi \lambda}U)$ is the wave packet with its retarded time rescaled, which for a field on a light front is a Lorentz boost along the propagation direction, and $\sigma$ is the symplectic form of Eq.~\eqref{eq:symplectic_form}. In the second line, $\langle :T_{UU}: \rangle_{\omega_\phi}$ is the normal-ordered energy-flux density of the wave packet, that is, the energy it carries per unit retarded time and per unit transverse area; it vanishes in the vacuum. Equation~\eqref{eq:Dorau_Srel} is written in natural units, $\hbar = c = 1$. The step from its first to its second line is given in Appendix~\ref{app:sigma_to_T}.

Integrating the energy-flux density over the cross section $\mathcal{S}$ gives the optical power of the wave packet at retarded time $U$,
\begin{equation}
P(U) = \int_{\mathcal{S}} \langle :T_{UU}: \rangle_{\omega_\phi}\,d\mathrm{vol}_{\mathcal{S}} .
\label{eq:P_of_U}
\end{equation}
With $\hbar$ restored, the quantum relative entropy $S^{\mathrm{rel}}(\omega_0 \parallel \omega_\phi)$ between the vacuum and the coherent state is then the integral over the retarded time $U$ of the power weighted by $|U|$,
\begin{equation}
S^{\mathrm{rel}}(\omega_0 \parallel \omega_\phi) = \frac{2\pi}{\hbar}\int |U|\,P(U)\,dU .
\label{eq:Srel_power}
\end{equation}
Equation~\eqref{eq:Srel_power} states that the relative entropy is the energy of the coherent wave packet in units of $\hbar$, with each part weighted by its retarded-time distance $|U|$ from the edge of the region. The weight $|U|$ arises because the vacuum restricted to a half space is a thermal state with respect to the boost generator~\cite{Bisognano1975}, and the relative entropy of a coherent state is the expectation value of that generator~\cite{Casini2019,Longo2019}. In the crystal, $\mathcal{S}$ is the cross section of the collinear mode and $P(U)$ is the power of the seed.

The optical counterpart of the coherent excitation is direct: the cw coherent seed $|\alpha\rangle$ is a Weyl displacement $\hat{W}(\alpha)$ of the idler vacuum that drives the nonlinear interaction. Because the QFT vacuum $\omega_0$ is entangled across causal partitions, displacing it by $\hat{W}(\alpha)$ adds a separable coherent component and lowers the relative weight of vacuum entanglement; $S^{\mathrm{rel}}(\omega_0 \parallel \omega_\alpha)$ quantifies this entanglement deficit~\cite{Dorau2026,Casini2019}.

\subsection{SPACS overlap fidelity}
\label{subsec:fidelity}

The same mechanism operates in the seeded ENBS interferometer of Fig.~\ref{fig:fig1}. Without the cw idler seed ($\alpha_j = 0$), spontaneous pair creation populates the orthogonal vacuum-fluctuation sectors $|1,0\rangle_s|1,0\rangle_{\delta i}$ and $|0,1\rangle_s|0,1\rangle_{\delta i}$~\cite{Yoon2026c}. In this case the SPACS of Eq.~\eqref{eq:SPACS} reduce to single idler photons in the two orthogonal idler modes, $\langle m_2 | m_1 \rangle_i = 0$, and there is no seed phase, so that the preparation phase reduces to the pump phase difference, $\Delta\phi_{\mathrm{prep}} \to \Delta\phi_p$. For symmetric pumping, $|c_1| = |c_2| = 1/\sqrt{2}$, the joint state of Eq.~\eqref{eq:joint_state} becomes
\begin{equation}
|\Psi_{si}\rangle = \frac{1}{\sqrt{2}}\Big(|1,0\rangle_s|1,0\rangle_{\delta i} + e^{-i\Delta\phi_p}\,|0,1\rangle_s|0,1\rangle_{\delta i}\Big).
\label{eq:unseeded_state}
\end{equation}
The SPFC signal supermode is then maximally entangled with the unmeasured idler modes ($E = 1$), and the first-order coherence vanishes ($V = 0$) for any value of $\Delta\phi_p$. With the shared seed $\hat{W}(\alpha_j)$ injected, stimulated emission populates the common zero-fluctuation sector $|0,0\rangle_{\delta i}$ with the seed photon number $|\alpha_j|^2 \propto \int P(U)\,dU$, where $P(U)$ is the seed power of Eq.~\eqref{eq:P_of_U}.

For symmetric seeding ($|\alpha_1| = |\alpha_2| = |\alpha|$), the overlap fidelity of the idler SPACS $|m_1\rangle_i$ and $|m_2\rangle_i$ after tracing over the shared seed reservoir is~\cite{Yoon2021,Yoon2026c}
\begin{align}
F(\alpha) = \big|\langle m_2 | m_1 \rangle_i\big| &= \frac{|\alpha|^2}{1 + |\alpha|^2} \nonumber\\
&= \frac{S^{\mathrm{rel}}(\omega_0 \parallel \omega_\alpha)}{s_1 + S^{\mathrm{rel}}(\omega_0 \parallel \omega_\alpha)}.
\label{eq:fidelity_Srel}
\end{align}
Here $S^{\mathrm{rel}}(\omega_0 \parallel \omega_\alpha) = s_1 |\alpha|^2$ is the relative entropy of the cw coherent seed within a window of duration $\tau_c$, and
\begin{equation}
s_1 \equiv \pi\,\omega_{\mathrm{sd}}\,\tau_c
\label{eq:s1}
\end{equation}
is the relative entropy per seed photon, a dimensionless number (Appendix~\ref{app:seed_srel}). The window $\tau_c$ is the duration of the idler temporal mode that the pump pulse couples to the signal supermode, and $r$ is the rate of seed photons in that mode, so that $|\alpha|^2 = r\,\tau_c$; it is not the coherence time of the seed laser. In Ref.~\cite{Lee2018} the mode bandwidth is about 50~GHz, which gives $\tau_c \approx 20~\mathrm{ps}$ and $s_1 \approx 8\times 10^{4}$, and the seed photon number was determined from the ratio of the stimulated and spontaneous down-conversion efficiencies, $|\alpha|^2 \approx 97$ at full seed power. Increasing the relative entropy of the seed therefore increases the entanglement deficit between the SPFC signal photon and the idler modes.

\subsection{Triality relation}
\label{subsec:triality}

\begin{figure*}[t]
\centering
\includegraphics[width=\textwidth]{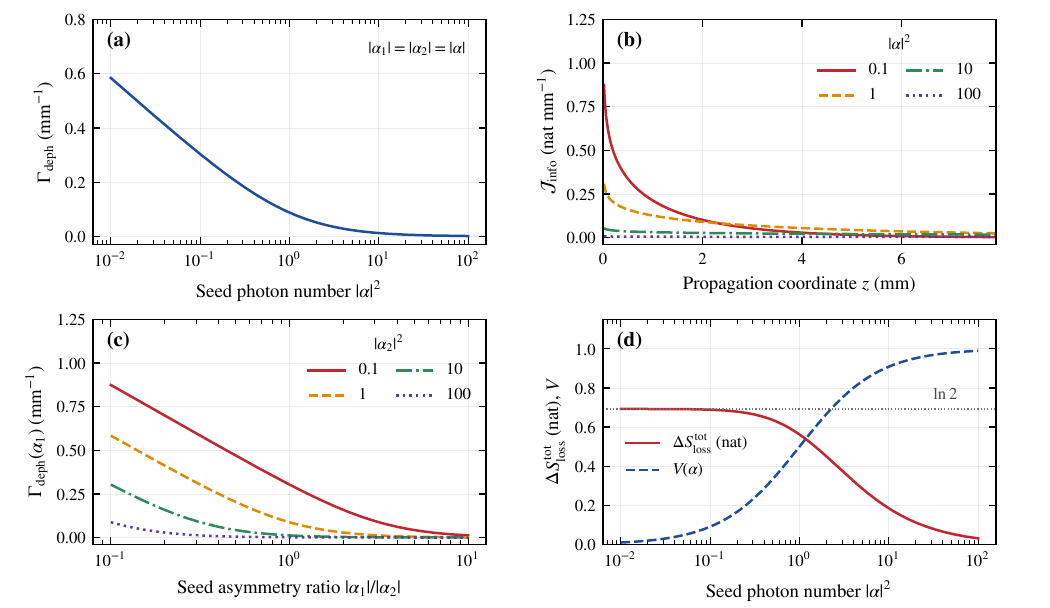}
\caption{Open-system dephasing and relative-entropy loss~\cite{Yoon2021,Yoon2026c,Yoon2026d}. (a) Spatial Lindblad dephasing rate $\Gamma_{\mathrm{deph}}$ of Eq.~\eqref{eq:gamma_deph} for symmetric seeding, which vanishes asymptotically as the seed photon number $|\alpha|^2$ increases. (b) Information loss rate $\mathcal{J}_{\mathrm{info}}(z)$ of Eq.~\eqref{eq:info_flux} along the propagation coordinate $z$; it decreases along the crystal, and its value near the crystal entrance is reduced by stronger seeding. (c) Local dephasing rate $\Gamma_{\mathrm{deph}}(\alpha_1)$ in the interaction volume $\mathcal{V}_1$ as a function of the seed asymmetry ratio $|\alpha_1|/|\alpha_2|$ for several values of the distant seed $|\alpha_2|^2$; the local rate, defined below Eq.~\eqref{eq:gamma_deph}, depends only on the local seed, consistent with Eq.~\eqref{eq:microcausality}. (d) Quantum-to-classical transition~\cite{Zavatta2004}: the integrated information loss $\Delta S_{\mathrm{loss}}^{\mathrm{tot}}$ of Eq.~\eqref{eq:total_info_loss} is bounded by $\ln 2$, its value for the unseeded source, and decays to zero as the visibility $V(\alpha)$ approaches unity.}
\label{fig:fig2}
\end{figure*}

With the path predictability $P = \big||c_1|^2 - |c_2|^2\big|$, the first-order interference visibility $V = 2|c_1 c_2|F(\alpha)$, and the signal-idler concurrence $E = 2|c_1 c_2|\sqrt{1 - F(\alpha)^2}$, the state of Eq.~\eqref{eq:joint_state} satisfies the wave-particle-entanglement triality identity~\cite{Jakob2010,Qian2020,Yoon2021}
\begin{equation}
P^2 + E^2 + V^2 = 1.
\label{eq:triality}
\end{equation}
Combining Eqs.~\eqref{eq:fidelity_Srel} and \eqref{eq:triality}, with $|\alpha|^2 = S^{\mathrm{rel}}/s_1$ and $S^{\mathrm{rel}} \equiv S^{\mathrm{rel}}(\omega_0 \parallel \omega_\alpha)$, gives the dependence of the entanglement on the seed and on its relative entropy,
\begin{align}
E(\alpha) &= 2|c_1 c_2|\,\frac{\sqrt{1+2|\alpha|^2}}{1+|\alpha|^2} \nonumber\\
&= 2|c_1 c_2|\,\frac{\sqrt{s_1\,(s_1 + 2S^{\mathrm{rel}})}}{s_1 + S^{\mathrm{rel}}} .
\label{eq:E_alpha}
\end{align}
The entanglement equals $2|c_1 c_2|$ for $S^{\mathrm{rel}} = 0$ and decreases monotonically with the relative entropy of the cw coherent seed, as $E \simeq 2|c_1 c_2|\sqrt{2 s_1/S^{\mathrm{rel}}}$ for $S^{\mathrm{rel}} \gg s_1$. The visibility increases correspondingly, $V = 2|c_1 c_2|\,S^{\mathrm{rel}}/(s_1 + S^{\mathrm{rel}})$. Bipartite vacuum entanglement is thus converted into single-photon coherence $V(\alpha)$ [Fig.~\ref{fig:fig2}(d)] without postselection and without overlapping the idler beams~\cite{Lee2018,Yoon2021,Yoon2026c}.

%=====================================================================
\section{Spatial Lindblad master equation and information loss rate}
\label{sec:lindblad}
%=====================================================================

\subsection{Master equation and dephasing rate}
\label{subsec:master}

In Ref.~\cite{Dorau2026} the variation of the relative entropy along the null coordinate $U$ defines the energy-information flux across the horizon. In the present system the corresponding evolution parameter is the propagation coordinate $z \in [0, L_c]$ inside the PPLN crystals (Fig.~\ref{fig:fig1}). Tracing out the unmeasured cw idler modes gives a spatial Lindblad master equation for the reduced SPFC signal density operator $\hat{\rho}_s(z) = \mathrm{Tr}_i\big(|\Psi_{si}(z)\rangle\langle\Psi_{si}(z)|\big)$~\cite{Lindblad1976,Carmichael1993,Yoon2021,Yoon2026d},
\begin{align}
\frac{d\hat{\rho}_s(z)}{dz} ={}& -\frac{i}{\hbar v_g}\big[\hat{H}_{\mathrm{eff}}, \hat{\rho}_s(z)\big] \nonumber\\
& + \frac{\Gamma_{\mathrm{deph}}}{2}\big(\hat{\tau}_z \hat{\rho}_s(z) \hat{\tau}_z - \hat{\rho}_s(z)\big),
\label{eq:spatial_lindblad}
\end{align}
where $v_g$ is the signal group velocity and $\hat{\tau}_z = |1,0\rangle_s\langle 1,0| - |0,1\rangle_s\langle 0,1|$ is the branch Pauli operator. With this normalization of the dissipator, which is that of Ref.~\cite{Yoon2026d}, $\Gamma_{\mathrm{deph}}$ is the decay rate of the off-diagonal elements of $\hat{\rho}_s$, and it is fixed by the SPACS overlap fidelity [Figs.~\ref{fig:fig2}(a) and \ref{fig:fig2}(c)],
\begin{equation}
\Gamma_{\mathrm{deph}}(\alpha) = -\frac{1}{L_c}\ln F(\alpha) = \frac{1}{L_c}\ln\!\left(1 + \frac{1}{|\alpha|^2}\right).
\label{eq:gamma_deph}
\end{equation}
Equation~\eqref{eq:gamma_deph} is written for symmetric seeding. For unequal seeds, the idler states of Eq.~\eqref{eq:SPACS} have the overlap found in Ref.~\cite{Lee2018},
\begin{equation}
\langle m_2 | m_1 \rangle_i = \frac{|\alpha_1||\alpha_2|}{\sqrt{(1+|\alpha_1|^2)(1+|\alpha_2|^2)}} = \sqrt{F(\alpha_1)\,F(\alpha_2)} ,
\label{eq:overlap_unequal}
\end{equation}
so that $-L_c^{-1}\ln\langle m_2 | m_1 \rangle_i = \tfrac{1}{2}[\Gamma_{\mathrm{deph}}(\alpha_1) + \Gamma_{\mathrm{deph}}(\alpha_2)]$, with the local rate $\Gamma_{\mathrm{deph}}(\alpha_j) = L_c^{-1}\ln(1 + 1/|\alpha_j|^2)$ of crystal $j$. The dephasing rate is thus the average of two local rates, each of which depends only on the seed of its own crystal. The local rate is the quantity labeled in Fig.~\ref{fig:fig1} and plotted in Fig.~\ref{fig:fig2}(c), and both local rates coincide with Eq.~\eqref{eq:gamma_deph} for symmetric seeding~\cite{Yoon2021,Yoon2026d}.

\subsection{Relative-entropy loss rate}
\label{subsec:flux}

In the branch basis $\{|1,0\rangle_s, |0,1\rangle_s\}$ and for symmetric pumping ($|c_1| = |c_2| = 1/\sqrt{2}$), the reduced signal density operator at position $z$ takes the form
\begin{equation}
\hat{\rho}_s(z) = \frac{1}{2}\begin{pmatrix} 1 & \mathcal{C}(z)\, e^{-i\Delta\Phi} \\ \mathcal{C}(z)\, e^{i\Delta\Phi} & 1 \end{pmatrix},
\label{eq:rho_z}
\end{equation}
where $\Delta\Phi$ is the total phase defined in Eq.~\eqref{eq:total_gauge_phase} below. The coherence factor follows from Eq.~\eqref{eq:spatial_lindblad}, which gives $d\mathcal{C}/dz = -\Gamma_{\mathrm{deph}}\,\mathcal{C}$ with $\mathcal{C}(0) = 1$. Its solution is $\mathcal{C}(z) = e^{-\Gamma_{\mathrm{deph}} z}$. Inserting the rate of Eq.~\eqref{eq:gamma_deph}, the coherence factor has the closed form
\begin{equation}
\mathcal{C}(z) = F(\alpha)^{z/L_c} = \left(\frac{|\alpha|^2}{1+|\alpha|^2}\right)^{z/L_c} .
\label{eq:coh_factor}
\end{equation}
It equals unity at the crystal entrance and reaches the SPACS overlap at the exit, $\mathcal{C}(L_c) = F(\alpha)$.

As a reference state we take the fully dephased branch mixture
\begin{equation}
\hat{\rho}_{\mathrm{mix}} = \frac{1}{2}\big(\hat{\rho}_s(z) + \hat{\tau}_z \hat{\rho}_s(z) \hat{\tau}_z\big) = \frac{1}{2}\hat{\mathbb{I}}_s .
\label{eq:rho_mix}
\end{equation}
This state is the same for every position $z$: the average removes the off-diagonal elements of Eq.~\eqref{eq:rho_z}, whatever the value of $\mathcal{C}(z)$, and the diagonal elements do not depend on $z$. The reference state is therefore fixed while $\hat{\rho}_s(z)$ evolves. Conjugation with $\hat{\tau}_z$ applies a $\pi$ phase shift to one supermode and reverses the sign of the off-diagonal elements, so the average in Eq.~\eqref{eq:rho_mix} removes the which-path coherence and leaves the diagonal classical mixture. The relative entropy of the signal state with respect to this reference is $S^{\mathrm{rel}}(\hat{\rho}_s \parallel \hat{\rho}_{\mathrm{mix}}) = \mathrm{Tr}[\hat{\rho}_s(\ln\hat{\rho}_s - \ln\hat{\rho}_{\mathrm{mix}})]$. Because $\ln\hat{\rho}_{\mathrm{mix}} = -\ln 2\;\hat{\mathbb{I}}_s$, it equals $\ln 2 + \sum_\pm \lambda_\pm \ln\lambda_\pm$, where $\lambda_\pm(z) = [1 \pm \mathcal{C}(z)]/2$ are the eigenvalues of $\hat{\rho}_s(z)$. Hence
\begin{equation}
S^{\mathrm{rel}}\big(\hat{\rho}_s(z) \parallel \hat{\rho}_{\mathrm{mix}}\big) = \ln 2 - H_{\mathrm{bin}}\!\left(\frac{1 + \mathcal{C}(z)}{2}\right),
\label{eq:Srel_z}
\end{equation}
where $H_{\mathrm{bin}}(x) = -x\ln x - (1-x)\ln(1-x)$ is the binary Shannon entropy. This quantity is the relative entropy of coherence of $\hat{\rho}_s(z)$ in the branch basis~\cite{Baumgratz2014}. In the following we write $S^{\mathrm{rel}}(z) \equiv S^{\mathrm{rel}}(\hat{\rho}_s(z) \parallel \hat{\rho}_{\mathrm{mix}})$. Writing out the binary entropy, the term $\ln 2$ cancels and the relative entropy takes the form
\begin{equation}
S^{\mathrm{rel}}(z) = \frac{1}{2}\big[(1+\mathcal{C})\ln(1+\mathcal{C}) + (1-\mathcal{C})\ln(1-\mathcal{C})\big],
\label{eq:Srel_C}
\end{equation}
with $\mathcal{C} \equiv \mathcal{C}(z)$. Inserting Eq.~\eqref{eq:coh_factor} into Eq.~\eqref{eq:Srel_C} gives the relative entropy of the present system explicitly as a function of the position and of the seed,
\begin{equation}
S^{\mathrm{rel}}(z) = \frac{1}{2}\sum_{\pm}\big(1 \pm F^{z/L_c}\big)\ln\!\big(1 \pm F^{z/L_c}\big),
\label{eq:Srel_explicit}
\end{equation}
where the sum runs over the two signs, as in the two terms of Eq.~\eqref{eq:Srel_C}, and $F = |\alpha|^2/(1+|\alpha|^2)$. It decreases monotonically from $S^{\mathrm{rel}}(0) = \ln 2$ at the entrance, where the signal state is pure, to
\begin{equation}
S^{\mathrm{rel}}(L_c) = \frac{1}{2}\big[(1+F)\ln(1+F) + (1-F)\ln(1-F)\big]
\label{eq:Srel_exit}
\end{equation}
at the exit. All entropies are defined with the natural logarithm, so they are dimensionless numbers expressed in nats, the natural unit of information ($1~\mathrm{nat} = 1/\ln 2 \approx 1.44$ bits).

We define the spatial information loss rate $\mathcal{J}_{\mathrm{info}}(z)$ as the relative entropy lost per unit length along the crystal [Fig.~\ref{fig:fig2}(b)]. Differentiating Eq.~\eqref{eq:Srel_C}, with $dS^{\mathrm{rel}}/d\mathcal{C} = \tfrac{1}{2}\ln[(1+\mathcal{C})/(1-\mathcal{C})] = \operatorname{artanh}\mathcal{C}$ and $d\mathcal{C}/dz = -\Gamma_{\mathrm{deph}}\,\mathcal{C}$, gives
\begin{align}
\mathcal{J}_{\mathrm{info}}(z) &\equiv -\frac{d S^{\mathrm{rel}}(z)}{dz} \nonumber\\
&= \Gamma_{\mathrm{deph}}(\alpha)\,\mathcal{G}_{\mathrm{coh}}\big(\hat{\rho}_s(z)\big) \ge 0,
\label{eq:info_flux}
\end{align}
with the non-negative coherence-divergence kernel
\begin{equation}
\mathcal{G}_{\mathrm{coh}}\big(\hat{\rho}_s(z)\big) = \mathcal{C}(z)\operatorname{artanh}\mathcal{C}(z) = \frac{\mathcal{C}(z)}{2}\ln\!\left(\frac{1 + \mathcal{C}(z)}{1 - \mathcal{C}(z)}\right).
\label{eq:Gcoh}
\end{equation}
With Eqs.~\eqref{eq:gamma_deph} and \eqref{eq:coh_factor}, the loss rate of the present system is, explicitly,
\begin{equation}
\mathcal{J}_{\mathrm{info}}(z) = \frac{1}{L_c}\ln\!\left(\frac{1}{F}\right) F^{z/L_c}\operatorname{artanh}\!\big(F^{z/L_c}\big).
\label{eq:J_explicit}
\end{equation}
Near the entrance, where $\mathcal{C} \simeq 1 - \Gamma_{\mathrm{deph}} z$, it diverges logarithmically, $\mathcal{J}_{\mathrm{info}} \simeq (\Gamma_{\mathrm{deph}}/2)\ln[2/(\Gamma_{\mathrm{deph}} z)]$, which is integrable, and at the exit it takes the finite value $L_c^{-1}\ln(1/F)\,F\operatorname{artanh}F$. The rate $\mathcal{J}_{\mathrm{info}}$ has the SI dimension of inverse length, $\mathrm{m}^{-1}$, like $\Gamma_{\mathrm{deph}}$: it is the relative entropy, in nats, that is lost per meter of propagation. The corresponding loss rate per unit time, for a signal wave packet moving at the group velocity $v_g$, is $v_g\,\mathcal{J}_{\mathrm{info}}$.

The total information transferred to the unmeasured idler modes during propagation [Fig.~\ref{fig:fig2}(d)] is the difference of Eq.~\eqref{eq:Srel_explicit} between the entrance and the exit,
\begin{align}
\Delta S_{\mathrm{loss}}^{\mathrm{tot}} &= \int_{0}^{L_c} \mathcal{J}_{\mathrm{info}}(z)\,dz = S^{\mathrm{rel}}(0) - S^{\mathrm{rel}}(L_c) \nonumber\\
&= H_{\mathrm{bin}}\!\left(\frac{1 + F(\alpha)}{2}\right),
\label{eq:total_info_loss}
\end{align}
where the last form follows from Eqs.~\eqref{eq:Srel_z} and \eqref{eq:Srel_exit}. It is bounded by $\ln 2$.

Equation~\eqref{eq:info_flux} is the open-system counterpart, for the collinear optical field, of the horizon flux law of Ref.~\cite{Dorau2026}. For $|\alpha| \to 0$ ($F \to 0$) the rate $\Gamma_{\mathrm{deph}}$ diverges and the off-diagonal coherence is lost completely, leaving the classical mixture $\hat{\rho}_{\mathrm{mix}}$ and $\Delta S_{\mathrm{loss}}^{\mathrm{tot}} = \ln 2$. For $|\alpha| \gg 1$ ($F \to 1$) the cw seed suppresses the loss rate ($\mathcal{J}_{\mathrm{info}} \to 0$) and the signal state remains pure over $L_c$.

%=====================================================================
\section{Quotient measurement projection and three-scan equivalence}
\label{sec:quotient}
%=====================================================================

\subsection{Tensor product as a quotient space}
\label{subsec:quotient_space}

A conceptual parallel between AQFT horizon theory~\cite{Dorau2026,Summers1996} and the dual-crystal ENBS~\cite{Yoon2026b,Yoon2026d} is that neither requires an instantaneous nonlocal collapse of the state. In textbook treatments, detecting a signal photon at the interferometer output is often said to collapse the distant idler field instantaneously. In Scholes' free-vector-space formulation of the tensor product~\cite{Scholes2026a,Scholes2026b,Scholes2026c,Yoon2026d}, the product space is the quotient
\begin{equation}
\mathcal{H}_s \otimes \mathcal{H}_i = F(\mathcal{H}_s \times \mathcal{H}_i)/\mathcal{N}
\label{eq:quotient_space}
\end{equation}
of the free vector space on formal pairs $(\psi_s, \psi_i)$ by the null subspace $\mathcal{N}$, which contains the combinations of pairs that must vanish for the pairs to behave as products. The construction is summarized, with examples for the present system, in Appendix~\ref{app:quotient}. A composite state such as $|\Psi_{si}\rangle$ is an equivalence class (coset) of sums of formal pairs. Within a coset, phase factors can be moved between the SPFC signal factor and the SPACS idler factor without changing their sign,
\begin{equation}
\big(e^{i\theta}\psi_s, \psi_i\big) \sim \big(\psi_s, e^{i\theta}\psi_i\big) \pmod{\mathcal{N}},
\label{eq:coset}
\end{equation}
because the tensor product is linear in each factor. A sign reversal occurs only when a state is used as a bra, as for the detector mode in Eq.~\eqref{eq:coset_slice} below.

\subsection{Local projection and coset slice}
\label{subsec:coset_slice}

When the two SPFC signal supermodes are recombined at the output beam splitter with post-creation interferometer phase $\Delta\phi_s$ (Fig.~\ref{fig:fig1}), the positive-frequency detection operator $\hat{E}_s^{(+)}(\Delta\phi_s) \propto (\hat{S}_1 + e^{i\Delta\phi_s}\hat{S}_2)/\sqrt{2}$ defines orthogonal bright and dark detector eigenmodes~\cite{VillasBoas2025,Yoon2026b},
\begin{align}
|B(\Delta\phi_s)\rangle_s &= \frac{1}{\sqrt{2}}\big(|1,0\rangle_s + e^{-i\Delta\phi_s}|0,1\rangle_s\big), \label{eq:bright_mode}\\
|D(\Delta\phi_s)\rangle_s &= \frac{1}{\sqrt{2}}\big(|1,0\rangle_s - e^{-i\Delta\phi_s}|0,1\rangle_s\big), \label{eq:dark_mode}
\end{align}
with $\hat{E}_s^{(+)}|D(\Delta\phi_s)\rangle_s = 0$. The bright-port projector
\begin{equation}
\hat{\Pi}_B(\Delta\phi_s) = |B(\Delta\phi_s)\rangle_s\langle B(\Delta\phi_s)| \otimes \hat{\mathbb{I}}_i
\label{eq:PiB}
\end{equation}
acts only on $\mathcal{H}_s$. Because the local field algebras commute at spacelike separation [Eq.~\eqref{eq:microcausality}], it does not disturb the distant idler fields, which are never recombined or detected. Contraction with the detector bra ${}_s\langle B(\Delta\phi_s)|$ instead selects the conditional SPACS coset slice~\cite{Yoon2026d}
\begin{align}
|\chi_i(\Delta\Phi)\rangle &= {}_s\langle B(\Delta\phi_s)|\Psi_{si}\rangle \nonumber\\
&= \frac{1}{\sqrt{2}}\left(c_1 |m_1\rangle_i + c_2 e^{i\Delta\Phi}|m_2\rangle_i\right),
\label{eq:coset_slice}
\end{align}
where the factor $e^{i\Delta\phi_s}$ from the detector bra multiplies the preparation phase $e^{-i\Delta\phi_{\mathrm{prep}}}$ of Eq.~\eqref{eq:joint_state}. Through the coset equivalence of Eq.~\eqref{eq:coset}, the readout phase $\Delta\phi_s$, the pump phase $\Delta\phi_p$, and the seed phase $\Delta\phi_{\mathrm{sd}}$ enter only in the combination
\begin{equation}
\Delta\Phi = \Delta\phi_s - \Delta\phi_p + \Delta\phi_{\mathrm{sd}}.
\label{eq:total_gauge_phase}
\end{equation}

\subsection{Counting rate and three-scan equivalence}
\label{subsec:three_scan}

The single-photon counting rate at the bright port is the squared norm of the projected coset slice,
\begin{align}
R_B(\Delta\Phi) &\propto \langle\chi_i(\Delta\Phi)|\chi_i(\Delta\Phi)\rangle \nonumber\\
&= \frac{1}{2}\big[1 + V(\alpha)\cos\Delta\Phi\big].
\label{eq:count_rate}
\end{align}
Equation~\eqref{eq:count_rate} accounts for the \emph{three-scan equivalence} observed experimentally~\cite{Lee2018,Yoon2026b,Yoon2026c} and discussed in Sec.~\ref{sec:comparison}. Scanning $\Delta\phi_p$ or $\Delta\phi_{\mathrm{sd}}$, which rotates the prepared state of the path qubit about its $\hat{\tau}_z$ axis, or scanning $\Delta\phi_s$, which rotates the readout basis about the same axis, traverses the same trajectory in the quotient space and gives identical fringes~\cite{Yoon2026d}. First-order single-photon interference thus arises from the local self-interference of the selected coset representative, with no dynamical action across the spacelike interval.

%=====================================================================
\section{Comparison with reported measurements}
\label{sec:comparison}
%=====================================================================

No new data are presented in this paper. The model can, however, be compared with the measurements of Ref.~\cite{Lee2018}, which were made with the setup described in Sec.~\ref{subsec:geometry}.

\subsection{Three-scan equivalence}
\label{subsec:cmp_three_scan}

Figure~2(a) of Ref.~\cite{Lee2018} shows the signal counting rate as a function of the displacements $\Delta x_p$, $\Delta x_s$, and $\Delta x_{\mathrm{seed}}$ of the pump, signal, and seed delay stages. The three fringes have periods of about $0.53$, $0.80$, and $1.55~\mu\mathrm{m}$, which equal the wavelengths of the scanned fields, $\lambda_p = 530~\mathrm{nm}$, $\lambda_s = 807~\mathrm{nm}$, and $\lambda_i = 1542~\mathrm{nm}$, and they have the same maximum counting rate. With $\Delta\phi_j = 2\pi\,\Delta x_j/\lambda_j$ for $j \in \{p, s, \mathrm{sd}\}$, this is the dependence of Eq.~\eqref{eq:count_rate} on the single phase $\Delta\Phi$ of Eq.~\eqref{eq:total_gauge_phase}. It agrees with the counting-rate formula of Ref.~\cite{Lee2018}, in which the pump and seed phase differences are defined with the opposite order of the two arms.

\subsection{Seed strength}
\label{subsec:cmp_strength}

Figure~3(b) of Ref.~\cite{Lee2018} shows the fringe visibility as a function of the seed photon number up to $|\alpha|^2 \approx 100$. The measured visibility rises steeply for $|\alpha|^2 \lesssim 10$ and saturates near $0.9$, consistent with
\begin{equation}
V_{\mathrm{exp}}(\alpha) = V_0\,F(\alpha) = V_0\,\frac{|\alpha|^2}{1+|\alpha|^2},
\label{eq:V_exp}
\end{equation}
with $V_0 \approx 0.9$, which is the function fitted to the data in Ref.~\cite{Lee2018}. The factor $V_0 \le 1$ accounts for the incomplete spatial overlap of the two signal beams at the output BS in that measurement~\cite{Lee2018}. It is not part of the ideal model, for which $V_0 = 1$, and it depends on the alignment.

Equation~\eqref{eq:V_exp} gives access to the quantities introduced in Sec.~\ref{sec:lindblad}. With $F = V_{\mathrm{exp}}/V_0$, the dephasing rate of Eq.~\eqref{eq:gamma_deph} and the information loss of Eq.~\eqref{eq:total_info_loss} follow from the measured visibility as
\begin{align}
\Gamma_{\mathrm{deph}} L_c &= -\ln\!\left(\frac{V_{\mathrm{exp}}}{V_0}\right), \label{eq:Gamma_from_V}\\
\Delta S_{\mathrm{loss}}^{\mathrm{tot}} &= H_{\mathrm{bin}}\!\left(\frac{1 + V_{\mathrm{exp}}/V_0}{2}\right). \label{eq:S_from_V}
\end{align}
For example, the model gives $\Gamma_{\mathrm{deph}} L_c \approx 0.41$ and $\Delta S_{\mathrm{loss}}^{\mathrm{tot}} \approx 0.45$ ($0.65\ln 2$) at $|\alpha|^2 = 2$, and $\Gamma_{\mathrm{deph}} L_c \approx 0.010$ and $\Delta S_{\mathrm{loss}}^{\mathrm{tot}} \approx 0.031$ at $|\alpha|^2 = 100$.

\subsection{Seed asymmetry}
\label{subsec:cmp_asymmetry}

Figure~2(b) of Ref.~\cite{Lee2018} shows the visibility as a function of the amplitude transmission coefficient $T$ of a variable attenuator in one seed arm, so that the two seed amplitudes are $\alpha$ and $T\alpha$. The stimulated emission probability of crystal $j$ is proportional to $1 + |\alpha_j|^2$~\cite{Lee2018}, so that $|c_j|^2 = (1+|\alpha_j|^2)/(2 + |\alpha_1|^2 + |\alpha_2|^2)$. With the overlap of Eq.~\eqref{eq:overlap_unequal}, the visibility $V = 2|c_1 c_2|\langle m_2 | m_1 \rangle_i$ becomes
\begin{equation}
V = \frac{2|\alpha_1||\alpha_2|}{2 + |\alpha_1|^2 + |\alpha_2|^2} = \frac{2|T||\alpha|^2}{2 + |\alpha|^2 + |T\alpha|^2},
\label{eq:V_unequal}
\end{equation}
which is the expression used in Ref.~\cite{Lee2018} to describe these data. It vanishes for $T \to 0$, where the idler states of the two crystals are orthogonal and the signal carries full which-path information, and it approaches unity for balanced strong seeding, as observed.

%=====================================================================
\section{Time-frequency GKP stabilizers from the symplectic Weyl algebra}
\label{sec:gkp}
%=====================================================================

The symplectic Weyl algebra $(\mathcal{W}, \sigma)$, with $\sigma$ given by Eq.~\eqref{eq:symplectic_form}, that is used in AQFT to evaluate the horizon relative entropy~\cite{Dorau2026} also underlies continuous-variable quantum error correction on the SPFC platform~\cite{Yoon2026a,Yoon2026c}. The single photon in the $N$-mode SPFC envelope ($N \approx 209$, OFC repetition period $T_{\mathrm{rep}} = 2\pi/\Omega_{\mathrm{rep}} = 4~\mathrm{ns}$) is described by its arrival-time operator $\hat{T}$~\cite{Maccone2020} and its angular-frequency operator $\hat{\omega}$~\cite{Fabre2020,Descamps2024,Yoon2026a}, which are canonically conjugate,
\begin{equation}
[\hat{T}, \hat{\omega}] = i .
\label{eq:TF_canonical}
\end{equation}
We use these two operators, in SI units, throughout. A time delay $\delta T$ and an angular-frequency shift $\delta\omega$, collected in the phase-space vector $\bm{\xi} = (\delta T, \delta\omega)$, are generated by the chronocyclic Weyl displacement operators
\begin{equation}
\hat{\mathcal{D}}_{\mathrm{TF}}(\delta T, \delta\omega) = \exp\!\big[i(\delta\omega\,\hat{T} - \delta T\,\hat{\omega})\big],
\label{eq:Weyl_TF}
\end{equation}
which obey the composition law of Eq.~\eqref{eq:Weyl_product}~\cite{Dorau2026,Yoon2026a},
\begin{equation}
\hat{\mathcal{D}}_{\mathrm{TF}}(\bm{\xi}_1)\hat{\mathcal{D}}_{\mathrm{TF}}(\bm{\xi}_2) = e^{-\frac{i}{2}\sigma_{\mathrm{TF}}(\bm{\xi}_1,\bm{\xi}_2)}\,\hat{\mathcal{D}}_{\mathrm{TF}}(\bm{\xi}_1 + \bm{\xi}_2),
\label{eq:Weyl_product_TF}
\end{equation}
with the time-frequency symplectic form
\begin{equation}
\sigma_{\mathrm{TF}}(\bm{\xi}_1,\bm{\xi}_2) = \delta T_1\, \delta\omega_2 - \delta\omega_1\, \delta T_2 ,
\label{eq:sigma_TF}
\end{equation}
the phase-space realization of Eq.~\eqref{eq:symplectic_form}. Here $\delta T_k$ and $\delta\omega_k$ are real numbers, the time delay and the angular-frequency shift of the $k$th displacement $\bm{\xi}_k = (\delta T_k, \delta\omega_k)$.

Two displacements commute when their symplectic phase is a multiple of $2\pi$. The TF-GKP stabilizer group of the comb is generated by a frequency shift by one tooth spacing and a time delay by two pulse periods~\cite{Gottesman2001,Fluhmann2019,Fabre2020,Yoon2026a},
\begin{align}
\hat{\mathcal{S}}_\omega &= \hat{\mathcal{D}}_{\mathrm{TF}}(0, \Omega_{\mathrm{rep}}), \qquad \hat{\mathcal{S}}_T = \hat{\mathcal{D}}_{\mathrm{TF}}(2T_{\mathrm{rep}}, 0), \nonumber\\
& [\hat{\mathcal{S}}_\omega, \hat{\mathcal{S}}_T] = 0,
\label{eq:GKP_stabilizers}
\end{align}
for which $|\sigma_{\mathrm{TF}}| = 2T_{\mathrm{rep}}\Omega_{\mathrm{rep}} = 4\pi$. The logical Pauli operators are a time delay by one pulse period and a frequency shift by half a tooth spacing,
\begin{equation}
\bar{Z}_{\mathrm{GKP}} = \hat{\mathcal{D}}_{\mathrm{TF}}(T_{\mathrm{rep}},0), \qquad \bar{X}_{\mathrm{GKP}} = \hat{\mathcal{D}}_{\mathrm{TF}}(0,\Omega_{\mathrm{rep}}/2),
\label{eq:GKP_logical}
\end{equation}
which anticommute because $|\sigma_{\mathrm{TF}}| = T_{\mathrm{rep}}\Omega_{\mathrm{rep}}/2 = \pi$.

These displacements are set by the comb itself. In frequency units, $\hat{\mathcal{S}}_\omega$ is a shift by $f_r = 250~\mathrm{MHz}$ and $\bar{X}_{\mathrm{GKP}}$ a shift by $f_r/2 = 125~\mathrm{MHz}$; $\hat{\mathcal{S}}_T$ is a delay by $2T_{\mathrm{rep}} = 8~\mathrm{ns}$ and $\bar{Z}_{\mathrm{GKP}}$ a delay by $T_{\mathrm{rep}} = 4~\mathrm{ns}$. The SPFC supermode of Eq.~\eqref{eq:SPFC_supermode}, a comb of teeth spaced by $\Omega_{\mathrm{rep}}$ and of pulses spaced by $T_{\mathrm{rep}}$, is invariant under both stabilizers and under $\bar{Z}_{\mathrm{GKP}}$, up to its finite spectral and temporal envelopes. The code lattice is thus fixed by the repetition rate of the comb, which is the sense in which the TF-GKP qubit is hardware native~\cite{Yoon2026a}.

Applying opposite electro-optic time delays $\pm\delta T/2$ and acousto-optic frequency shifts $\pm\delta\omega/2$ in the two arms of the ENBS interferometer couples the path qubit $\mathcal{H}_T$ to the continuous SPFC grid mode through the controlled Weyl gate~\cite{Yoon2026c}
\begin{align}
\hat{C\mathcal{D}}(\delta T, \delta\omega) ={}& |+\rangle\langle +| \otimes \hat{\mathcal{D}}_{\mathrm{TF}}\!\left(+\frac{\delta T}{2}, +\frac{\delta\omega}{2}\right) \nonumber\\
& + |-\rangle\langle -| \otimes \hat{\mathcal{D}}_{\mathrm{TF}}\!\left(-\frac{\delta T}{2}, -\frac{\delta\omega}{2}\right),
\label{eq:controlled_displacement}
\end{align}
where the branch projectors $|+\rangle\langle+|$ and $|-\rangle\langle-|$ correspond to the two signal paths $|1,0\rangle_s$ and $|0,1\rangle_s$ of Eq.~\eqref{eq:joint_state}.

The signal photon thus carries two degrees of freedom: the path qubit $\mathcal{H}_T$ and its time-frequency degree of freedom, on which $\hat{T}$ and $\hat{\omega}$ act and whose Hilbert space we denote by $\mathcal{H}_{\mathrm{grid}}$. Let $\hat{\rho}_{\mathrm{grid}}$ be the density operator of the photon on $\mathcal{H}_{\mathrm{grid}}$, which is the pure state of the SPFC supermode of Eq.~\eqref{eq:SPFC_supermode} for the present source and an encoded TF-GKP state in general, and let $\hat{\mathcal{I}}_{\mathrm{grid}}$ be the identity operator on $\mathcal{H}_{\mathrm{grid}}$. We write $\langle\,\cdot\,\rangle_{\mathrm{out}}$ for the expectation value in the state of the signal photon after the gate of Eq.~\eqref{eq:controlled_displacement}, with the idler modes traced out. For symmetric pumping and $\Delta\Phi = 0$, the normalized difference of the bright- and dark-port counting rates of Eq.~\eqref{eq:count_rate} measures $\langle \hat{\tau}_x \otimes \hat{\mathcal{I}}_{\mathrm{grid}} \rangle_{\mathrm{out}}$. Evaluated at $(\delta T, \delta\omega) = (0, \Omega_{\mathrm{rep}})$ or $(2T_{\mathrm{rep}}, 0)$, it gives the stabilizer expectation values,
\begin{equation}
\big\langle \hat{\tau}_x \otimes \hat{\mathcal{I}}_{\mathrm{grid}} \big\rangle_{\mathrm{out}} = F(\alpha)\,\mathrm{Re}\,\mathrm{Tr}\!\left[\hat{\rho}_{\mathrm{grid}}\,\hat{\mathcal{D}}_{\mathrm{TF}}(\delta T,\delta\omega)\right],
\label{eq:syndrome_readout}
\end{equation}
where $\hat{\tau}_x = |1,0\rangle_s\langle 0,1| + |0,1\rangle_s\langle 1,0|$ is the branch Pauli operator conjugate to $\hat{\tau}_z$ of Eq.~\eqref{eq:spatial_lindblad}.
The contrast of the TF-GKP syndrome readout is therefore reduced by the factor $F(\alpha)$, with deficit $1 - F(\alpha) = (1 + |\alpha|^2)^{-1}$ set by the relative entropy $S^{\mathrm{rel}}(\omega_0 \parallel \omega_\alpha)$ of the cw coherent seed. This links the AQFT information measure to a practical figure of merit of bosonic error correction.

%=====================================================================
\section{Discussion and conclusion}
\label{sec:conclusion}
%=====================================================================

We have carried the relative-entropy structure used in the AQFT derivation of the semiclassical Einstein equations on local horizons~\cite{Dorau2026} over to a photonic system, the dual-crystal, comb-pumped, coherently seeded biphoton interferometer. A single quantity, the SPACS overlap $F(\alpha)$, organizes the description. It is a function of the relative entropy of the coherent seed [Eq.~\eqref{eq:fidelity_Srel}]; it fixes the visibility and the entanglement in the triality relation [Eq.~\eqref{eq:triality}]; it sets the rate of the spatial Lindblad dephasing [Eq.~\eqref{eq:gamma_deph}] and the total relative entropy transferred to the idler modes [Eq.~\eqref{eq:total_info_loss}]; and it limits the contrast of the TF-GKP stabilizer readout [Eq.~\eqref{eq:syndrome_readout}].

Within this model the observed single-photon interference requires no nonlocal collapse postulate. The detector at the signal output selects a coset representative of a composite state whose correlations were fixed when the pair was prepared, and the local algebras of the two crystals commute. The pump, seed, and readout phase scans are equivalent because they enter the coset slice only through the invariant phase $\Delta\Phi$.

Two limits of the analysis should be stated. First, it describes ensemble counting rates and reduced density operators; it does not address how a single outcome is actualized in an individual detection event, and it is not a test of local hidden-variable models, since first-order interference of the seeded source involves no Bell inequality. Second, the correspondence with the horizon flux law of Ref.~\cite{Dorau2026} is structural: the propagation coordinate $z$ plays the role of the affine parameter, and the reference state is the dephased mixture and not the vacuum. Quantitative tests are nevertheless available with existing sources~\cite{Lee2018,Yoon2021}: the seed-power dependence of the visibility and of the dephasing rate in Figs.~\ref{fig:fig2}(a) and \ref{fig:fig2}(d), the independence of the local rate from the distant seed in Fig.~\ref{fig:fig2}(c), and the stabilizer contrast of Eq.~\eqref{eq:syndrome_readout}.

\section*{Data availability}
No experimental data were created or analyzed in this study. The curves in Fig.~\ref{fig:fig2} are evaluated directly from the analytical expressions given in the text.

\begin{acknowledgments}
The author thanks S. K. Lee and M. Cho for their long-term collaboration on SPFC interferometry. This work was supported by the National Research Foundation of Korea (NRF) under Grant No.~RS-2022-NR068815, and in part by NRF Grant No.~RS-2023-NR068116 and by the Institute for Basic Science (IBS-R023-D1). An AI language model (Claude, Anthropic) was used for assistance with language editing and formatting of the manuscript; the author is responsible for all content.
\end{acknowledgments}

\appendix

%=====================================================================
\section{Dictionary between AQFT and quantum optics}
\label{app:AQFT_QED_bridge}
%=====================================================================

This Appendix summarizes the correspondences between the AQFT language of Dorau and Much~\cite{Dorau2026} and the operational formalism of quantum optics used in the main text.

\subsection{Divergent von Neumann entropy and finite relative entropy}
\label{app:type_III}

In discrete-mode quantum mechanics (a two-level system or a single cavity mode), the observable algebra is a type-I von Neumann algebra. It admits a Hilbert-space trace $\mathrm{Tr}(\cdot)$, a trace-class reduced density operator $\hat{\rho}$ with $\mathrm{Tr}\,\hat{\rho} = 1$, and a finite von Neumann entropy $S_{\mathrm{vN}}(\hat{\rho}) = -\mathrm{Tr}(\hat{\rho}\ln\hat{\rho})$. When a continuous multimode QFT is restricted to a bounded spacetime region $\mathcal{U}$, infinitely many ultraviolet vacuum modes are entangled across the boundary. In AQFT the local algebra is then a type-$\mathrm{III}_1$ von Neumann algebra~\cite{Araki1976,Summers1996,Dorau2026}, for which neither a trace nor a local reduced density operator exists. States are described as expectation functionals $\omega(\cdot)$. For a local observable $\hat{\mathcal{O}}$, the coherent-excitation functional
\begin{equation}
\omega_\phi(\hat{\mathcal{O}}) = \omega_0\big(\hat{W}(\phi)^\dagger \hat{\mathcal{O}}\, \hat{W}(\phi)\big)
\label{eq:app_functional}
\end{equation}
corresponds to the Hilbert-space expectation value $\langle \alpha | \hat{\mathcal{O}} | \alpha \rangle$ in a coherent state generated by a Weyl displacement.

\subsection{Entanglement deficit}
\label{app:deficit}

In AQFT the vacuum state $\omega_0$ is entangled across causal horizons. Acting with a local Weyl unitary $\hat{W}(\phi)$ creates a coherent wave packet on top of the vacuum and reduces the relative weight of vacuum entanglement per generated quantum~\cite{Dorau2026,Casini2019}. In the dual-crystal ENBS interferometer (Fig.~\ref{fig:fig1}), unseeded spontaneous down-conversion ($\alpha = 0$) generates signal-idler pairs from vacuum fluctuations alone, and for symmetric pumping the SPFC signal path is maximally entangled with orthogonal idler modes ($E = 1$, $V = 0$). The cw coherent seed $|\alpha\rangle$ stimulates pair emission into SPACS that do not distinguish which crystal emitted the photon. The AQFT entanglement deficit thus corresponds to the suppression of which-path entanglement, $E(\alpha) = \sqrt{1+2|\alpha|^2}/(1+|\alpha|^2)$ [Eq.~\eqref{eq:E_alpha}], in favor of the first-order visibility $V(\alpha) = |\alpha|^2/(1+|\alpha|^2)$.

\subsection{Continuous variables}
\label{app:variables}

On the horizon side, AQFT uses the affine retarded null coordinate $U = t - z/c$, symplectic test-function configurations $\phi$ that depend on $U$ and on the position on the horizon cross section $\mathcal{S}$, and the normal-ordered null energy-momentum tensor $:T_{UU}:$, for which $\langle :T_{UU}: \rangle_{\omega_0} = 0$ in the vacuum. The affine dilation $\phi^\lambda(U) = \phi(e^{2\pi \lambda}U)$ is a longitudinal Lorentz boost with rapidity parameter $\lambda$. Its symplectic derivative with respect to $\lambda$ generates the energy-momentum flux across $\mathcal{H}_B^{\mathcal{R}}$ that determines the relative entropy $S^{\mathrm{rel}}(\omega_0 \parallel \omega_\phi)$ of Eq.~\eqref{eq:Dorau_Srel}.

On the quantum-optics side, the corresponding variables are the arrival-time and angular-frequency operators $\hat{T}$ and $\hat{\omega}$ of Eq.~\eqref{eq:TF_canonical} on the $N$-mode comb envelope, the chronocyclic Weyl displacements $\hat{\mathcal{D}}_{\mathrm{TF}}(\delta T, \delta\omega)$ of Eq.~\eqref{eq:Weyl_TF}, and the propagation coordinate $z \in [0, L_c]$ along the PPLN crystals. The flow of the boost generator is replaced by evolution along the crystal axis, which drives the information loss rate $\mathcal{J}_{\mathrm{info}}(z)$ of Eq.~\eqref{eq:info_flux}.

\subsection{From the symplectic form to the energy flux}
\label{app:sigma_to_T}

The second line of Eq.~\eqref{eq:Dorau_Srel} follows from the first in three steps. We write $\phi' \equiv \partial_U\phi$ and consider one point of the cross section $\mathcal{S}$; the integral over $\mathcal{S}$ is restored at the end.

(i) Differentiating the rescaled wave packet $\phi^\lambda(U) = \phi(e^{2\pi\lambda}U)$ with respect to $\lambda$ gives
\begin{equation}
\frac{d\phi^\lambda(U)}{d\lambda}\Big|_{\lambda=0} = 2\pi\,U\,\phi'(U).
\label{eq:app_dphi}
\end{equation}

(ii) Because the symplectic form of Eq.~\eqref{eq:symplectic_form} is bilinear,
\begin{align}
\frac{1}{2}\frac{d}{d\lambda}\Big|_{\lambda=0}\sigma(\phi^\lambda,\phi) &= \pi\,\sigma(U\phi', \phi) \nonumber\\
&= \pi\int \big[U\phi'^2 - \phi\,\partial_U(U\phi')\big]\,dU .
\label{eq:app_sigma_step}
\end{align}
For a wave packet that vanishes at the ends of the integration range, an integration by parts gives $-\int \phi\,\partial_U(U\phi')\,dU = \int U\phi'^2\,dU$, so that
\begin{equation}
\frac{1}{2}\frac{d}{d\lambda}\Big|_{\lambda=0}\sigma(\phi^\lambda,\phi) = 2\pi\int U\,(\partial_U\phi)^2\,dU .
\label{eq:app_sigma_result}
\end{equation}

(iii) In the coherent state $\hat{W}(\phi)|0\rangle$ the field has a classical expectation value fixed by $\phi$, and in the normalization in which $[\hat{\Phi}(\phi), \hat{\Phi}(\psi)] = i\sigma(\phi,\psi)$ its normal-ordered energy-flux density is~\cite{Dorau2026,Casini2019,Longo2019}
\begin{equation}
\langle :T_{UU}: \rangle_{\omega_\phi} = (\partial_U\phi)^2 .
\label{eq:app_TUU}
\end{equation}
Inserting Eq.~\eqref{eq:app_TUU} into Eq.~\eqref{eq:app_sigma_result}, with $U = |U|$ on the part of the sheet where $U > 0$, and integrating over the cross section $\mathcal{S}$ gives the second line of Eq.~\eqref{eq:Dorau_Srel}.

%=====================================================================
\section{Relative entropy of the coherent seed}
\label{app:seed_srel}
%=====================================================================

To evaluate the relative entropy $S^{\mathrm{rel}}(\omega_0 \parallel \omega_\alpha)$ of the cw coherent seed that enters Eq.~\eqref{eq:fidelity_Srel}, we use the form of Eq.~\eqref{eq:Srel_power}, in which the energy-flux density has been integrated over the cross section of the collinear mode. For a monochromatic cw seed with angular frequency $\omega_{\mathrm{sd}}$ and photon rate $r$ (number of photons passing through the cross section of the collinear mode per unit time), the power is constant,
\begin{equation}
P = \hbar\,\omega_{\mathrm{sd}}\,r .
\label{eq:app_power}
\end{equation}
We consider the part of the seed inside the window $0 \le U \le \tau_c$ of the idler temporal mode defined below Eq.~\eqref{eq:s1}, taken to start at the edge $U = 0$ of the region. It contains $|\alpha|^2 = r\,\tau_c$ photons. Equation~\eqref{eq:Srel_power} then gives
\begin{align}
S^{\mathrm{rel}}(\omega_0 \parallel \omega_\alpha) &= \frac{2\pi}{\hbar}\int_0^{\tau_c} U\,\hbar\,\omega_{\mathrm{sd}}\,r\,dU \nonumber\\
&= \pi\,\omega_{\mathrm{sd}}\,r\,\tau_c^2 = \pi\,\omega_{\mathrm{sd}}\,\tau_c\,|\alpha|^2 .
\label{eq:app_seed_srel}
\end{align}
The result is dimensionless and linear in the seed photon number, with the coefficient $s_1 = \pi\,\omega_{\mathrm{sd}}\tau_c$ of Eq.~\eqref{eq:s1}, which is $\pi$ times the optical phase accumulated by the seed over the window. If the window starts at $U_0 > 0$ instead of at the edge, the coefficient becomes $2\pi\,\omega_{\mathrm{sd}}(U_0 + \tau_c/2)$. The relative entropy remains linear in $|\alpha|^2$, so the form of Eq.~\eqref{eq:fidelity_Srel} does not depend on this choice.

%=====================================================================
\section{Quotient construction of the tensor product}
\label{app:quotient}
%=====================================================================

This Appendix summarizes the construction behind Eq.~\eqref{eq:quotient_space}~\cite{Scholes2026a,Scholes2026b,Scholes2026c} and illustrates it with the seeded joint state of Eq.~\eqref{eq:joint_state}.

\subsection{Free vector space and null subspace}
\label{app:free_space}

The free vector space $F(\mathcal{H}_s \times \mathcal{H}_i)$ consists of all finite formal sums of ordered pairs $(\psi_s, \psi_i)$ with complex coefficients. Every pair is an independent basis element. For example, $(2\psi_s, \psi_i)$, $(\psi_s, 2\psi_i)$, and $2(\psi_s, \psi_i)$ are three unrelated elements, and so are $(e^{i\theta}\psi_s, \psi_i)$ and $(\psi_s, e^{i\theta}\psi_i)$. This space is much larger than the space of physical states, because it does not yet know that a pair is meant to be a product.

The null subspace $\mathcal{N}$ is spanned by the combinations that must vanish for the pairs to behave as products,
\begin{align}
&(\psi_s + \psi_s', \psi_i) - (\psi_s, \psi_i) - (\psi_s', \psi_i), \nonumber\\
&(\psi_s, \psi_i + \psi_i') - (\psi_s, \psi_i) - (\psi_s, \psi_i'), \nonumber\\
&(c\,\psi_s, \psi_i) - c\,(\psi_s, \psi_i), \nonumber\\
&(\psi_s, c\,\psi_i) - c\,(\psi_s, \psi_i),
\label{eq:null_space}
\end{align}
for all states and all complex numbers $c$. The elements of $\mathcal{N}$ are not physical states. They are the formal expressions that represent the zero vector of the product space.

\subsection{Quotient as an identification}
\label{app:identification}

The quotient $F/\mathcal{N}$ is not a division of numbers but an identification. Two formal sums $x$ and $y$ are regarded as the same element whenever their difference $x - y$ lies in $\mathcal{N}$, in the same way as two potentials that differ by a gauge transformation describe the same field. The elements of the quotient are the equivalence classes (cosets) $x + \mathcal{N}$. They inherit addition and multiplication by complex numbers from the free vector space, so they form a vector space.

The equivalence class of the pair $(\psi_s, \psi_i)$ is the product state $\psi_s \otimes \psi_i$. It is bilinear because every expression in Eq.~\eqref{eq:null_space} is identified with zero. If $\{e_a\}$ and $\{f_b\}$ are bases of $\mathcal{H}_s$ and $\mathcal{H}_i$, the classes $e_a \otimes f_b$ form a basis of the quotient, whose dimension is therefore the product of the two dimensions. With the inner product
\begin{equation}
\langle \psi_s \otimes \psi_i | \psi_s' \otimes \psi_i' \rangle = \langle \psi_s | \psi_s' \rangle \langle \psi_i | \psi_i' \rangle ,
\label{eq:app_inner}
\end{equation}
extended by linearity, the quotient is the Hilbert space of the composite system. The spaces $\mathcal{H}_s$ and $\mathcal{H}_i$ are factors of this space and not subspaces of it.

For the present system the relevant signal space is spanned by the two supermode states $|1,0\rangle_s$ and $|0,1\rangle_s$, and the relevant idler space by the two SPACS $|m_1\rangle_i$ and $|m_2\rangle_i$, which are linearly independent for $F < 1$ although they are not orthogonal. The free vector space contains one basis element for every pair of vectors of these two spaces, a continuum, whereas the quotient has dimension four, with basis $|1,0\rangle_s|m_1\rangle_i$, $|1,0\rangle_s|m_2\rangle_i$, $|0,1\rangle_s|m_1\rangle_i$, and $|0,1\rangle_s|m_2\rangle_i$.

\subsection{The seeded joint state as a coset}
\label{app:joint_coset}

The joint state of Eq.~\eqref{eq:joint_state} is the coset of the formal sum
\begin{equation}
x_1 = c_1\big(|1,0\rangle_s, |m_1\rangle_i\big) + c_2\big(e^{-i\Delta\phi_{\mathrm{prep}}}|0,1\rangle_s, |m_2\rangle_i\big).
\label{eq:app_x1}
\end{equation}
Taking $c = e^{-i\Delta\phi_{\mathrm{prep}}}$ in the last two lines of Eq.~\eqref{eq:null_space} shows that
\begin{equation}
x_2 = c_1\big(|1,0\rangle_s, |m_1\rangle_i\big) + c_2\big(|0,1\rangle_s, e^{-i\Delta\phi_{\mathrm{prep}}}|m_2\rangle_i\big)
\label{eq:app_x2}
\end{equation}
differs from $x_1$ by an element of $\mathcal{N}$, so that $x_1$ and $x_2$ are two representatives of the same state. The same holds for
\begin{equation}
x_3 = c_1\big(|1,0\rangle_s, |m_1\rangle_i\big) + c_2\big(e^{-i\Delta\phi_p}|0,1\rangle_s, e^{i\Delta\phi_{\mathrm{sd}}}|m_2\rangle_i\big),
\label{eq:app_x3}
\end{equation}
in which the pump phase is attached to the signal factor and the seed phase to the idler factor, since $\Delta\phi_{\mathrm{prep}} = \Delta\phi_p - \Delta\phi_{\mathrm{sd}}$. The state does not record on which photon of the second crystal the preparation phase resides. This is the content of Eq.~\eqref{eq:coset}, and it is the algebraic reason why a scan of the pump phase and a scan of the seed phase are indistinguishable at the signal detector.

Three sign rules follow. First, a phase factor moved between the two factors of one pair keeps its sign, as in $x_1$ and $x_2$. Second, opposite phases on the two factors of one pair cancel, $(e^{i\theta}\psi_s, e^{-i\theta}\psi_i) \sim (\psi_s, \psi_i)$. Third, the relative phase between the two terms can be moved from the second term to the first only with the opposite sign and up to a global phase,
\begin{align}
& c_1|1,0\rangle_s|m_1\rangle_i + c_2 e^{-i\Delta\phi_{\mathrm{prep}}}|0,1\rangle_s|m_2\rangle_i \nonumber\\
&\quad = e^{-i\Delta\phi_{\mathrm{prep}}}\big(c_1 e^{i\Delta\phi_{\mathrm{prep}}}|1,0\rangle_s|m_1\rangle_i + c_2|0,1\rangle_s|m_2\rangle_i\big).
\label{eq:app_global}
\end{align}
This last step is not an identification by $\mathcal{N}$. It is the usual statement that a global phase is unobservable.

\subsection{Projection on the signal factor}
\label{app:projection}

The contraction with the detector bra in Eq.~\eqref{eq:coset_slice} acts on each pair as
\begin{equation}
(\psi_s, \psi_i) \mapsto {}_s\langle B(\Delta\phi_s)|\psi_s\rangle\,\psi_i ,
\label{eq:app_contraction}
\end{equation}
extended by linearity to formal sums. This map is linear in each entry of the pair, so it sends every expression in Eq.~\eqref{eq:null_space} to zero, and its result is the same for all representatives of a coset. Applied to $x_1$, $x_2$, or $x_3$ it gives
\begin{equation}
|\chi_i(\Delta\Phi)\rangle = \frac{1}{\sqrt{2}}\big(c_1|m_1\rangle_i + c_2 e^{i(\Delta\phi_s - \Delta\phi_{\mathrm{prep}})}|m_2\rangle_i\big),
\label{eq:app_slice}
\end{equation}
which is Eq.~\eqref{eq:coset_slice} with $\Delta\Phi = \Delta\phi_s - \Delta\phi_p + \Delta\phi_{\mathrm{sd}}$. The factor $e^{i\Delta\phi_s}$ has the sign opposite to that in the bright mode of Eq.~\eqref{eq:bright_mode} because the mode enters as a bra. The projection acts on the signal factor alone, and the idler factor of every pair is left unchanged.

\bibliography{Manuscript}

@article{Dorau2026,
  author  = {Dorau, Philipp and Much, Albert},
  title   = {From Quantum Relative Entropy to the Semiclassical {E}instein Equations},
  journal = {Phys. Rev. Lett.},
  volume  = {136},
  pages   = {091602},
  year    = {2026},
  doi     = {10.1103/lmq8-nsty}
}

@article{Jacobson1995,
  author  = {Jacobson, Ted},
  title   = {Thermodynamics of Spacetime: The {E}instein Equation of State},
  journal = {Phys. Rev. Lett.},
  volume  = {75},
  pages   = {1260--1263},
  year    = {1995},
  doi     = {10.1103/PhysRevLett.75.1260}
}

@article{Araki1976,
  author  = {Araki, Huzihiro},
  title   = {Relative Entropy of States of von {N}eumann Algebras},
  journal = {Publ. Res. Inst. Math. Sci.},
  volume  = {11},
  pages   = {809--833},
  year    = {1976},
  doi     = {10.2977/prims/1195191148}
}

@article{Uhlmann1977,
  author  = {Uhlmann, Armin},
  title   = {Relative Entropy and the {W}igner-{Y}anase-{D}yson-{L}ieb Concavity in an Interpolation Theory},
  journal = {Commun. Math. Phys.},
  volume  = {54},
  pages   = {21--32},
  year    = {1977},
  doi     = {10.1007/BF01609834}
}

@article{Casini2008,
  author  = {Casini, Horacio},
  title   = {Relative entropy and the {B}ekenstein bound},
  journal = {Class. Quantum Grav.},
  volume  = {25},
  pages   = {205021},
  year    = {2008},
  doi     = {10.1088/0264-9381/25/20/205021}
}

@article{Casini2019,
  author  = {Casini, Horacio and Grillo, Sergio and Pontello, Diego},
  title   = {Relative entropy for coherent states from {A}raki formula},
  journal = {Phys. Rev. D},
  volume  = {99},
  pages   = {125020},
  year    = {2019},
  doi     = {10.1103/PhysRevD.99.125020}
}

@article{Longo2019,
  author  = {Longo, Roberto},
  title   = {Entropy of Coherent Excitations},
  journal = {Lett. Math. Phys.},
  volume  = {109},
  pages   = {2587--2600},
  year    = {2019},
  doi     = {10.1007/s11005-019-01196-6}
}

@article{Hollands2019,
  author  = {Hollands, Stefan and Ishibashi, Akihiro},
  title   = {News versus information},
  journal = {Class. Quantum Grav.},
  volume  = {36},
  pages   = {195001},
  year    = {2019},
  doi     = {10.1088/1361-6382/ab3c1e}
}

@article{Kurpicz2021,
  author  = {Kurpicz, F. and Pinamonti, N. and Verch, R.},
  title   = {Temperature and entropy-area relation of quantum matter near spherically symmetric outer trapping horizons},
  journal = {Lett. Math. Phys.},
  volume  = {111},
  pages   = {110},
  year    = {2021},
  doi     = {10.1007/s11005-021-01445-7}
}

@article{Kay1991,
  author  = {Kay, Bernard S. and Wald, Robert M.},
  title   = {Theorems on the uniqueness and thermal properties of stationary, nonsingular, quasifree states on spacetimes with a bifurcate {K}illing horizon},
  journal = {Phys. Rep.},
  volume  = {207},
  pages   = {49--136},
  year    = {1991},
  doi     = {10.1016/0370-1573(91)90015-E}
}

@article{Summers1996,
  author  = {Summers, Stephen J. and Verch, Rainer},
  title   = {Modular inclusion, the {H}awking temperature, and quantum field theory in curved spacetime},
  journal = {Lett. Math. Phys.},
  volume  = {37},
  pages   = {145--158},
  year    = {1996},
  doi     = {10.1007/BF00416017}
}

@article{Bohr1928,
  author  = {Bohr, Niels},
  title   = {The Quantum Postulate and the Recent Development of Atomic Theory},
  journal = {Nature},
  volume  = {121},
  pages   = {580--590},
  year    = {1928},
  doi     = {10.1038/121580a0}
}

@article{Bell1964,
  author  = {Bell, John S.},
  title   = {On the {E}instein {P}odolsky {R}osen paradox},
  journal = {Physics Physique Fizika},
  volume  = {1},
  pages   = {195--200},
  year    = {1964},
  doi     = {10.1103/PhysicsPhysiqueFizika.1.195}
}

@article{Brunner2014,
  author  = {Brunner, Nicolas and Cavalcanti, Daniel and Pironio, Stefano and Scarani, Valerio and Wehner, Stephanie},
  title   = {{B}ell nonlocality},
  journal = {Rev. Mod. Phys.},
  volume  = {86},
  pages   = {419--478},
  year    = {2014},
  doi     = {10.1103/RevModPhys.86.419}
}

@book{Wiseman2010,
  author    = {Wiseman, Howard M. and Milburn, Gerard J.},
  title     = {Quantum Measurement and Control},
  publisher = {Cambridge University Press},
  address   = {Cambridge},
  year      = {2010}
}

@article{Zou1991,
  author  = {Zou, X. Y. and Wang, L. J. and Mandel, L.},
  title   = {Induced coherence and indistinguishability in optical interference},
  journal = {Phys. Rev. Lett.},
  volume  = {67},
  pages   = {318--321},
  year    = {1991},
  doi     = {10.1103/PhysRevLett.67.318}
}

@article{Wang1991,
  author  = {Wang, L. J. and Zou, X. Y. and Mandel, L.},
  title   = {Induced coherence without induced emission},
  journal = {Phys. Rev. A},
  volume  = {44},
  pages   = {4614--4622},
  year    = {1991},
  doi     = {10.1103/PhysRevA.44.4614}
}

@article{Lee2018,
  author  = {Lee, Sun Kyung and Han, Noh Soo and Yoon, Tai Hyun and Cho, Minhaeng},
  title   = {Frequency comb single-photon interferometry},
  journal = {Commun. Phys.},
  volume  = {1},
  pages   = {51},
  year    = {2018},
  doi     = {10.1038/s42005-018-0051-2}
}

@article{Yoon2021,
  author  = {Yoon, Tai Hyun and Cho, Minhaeng},
  title   = {Quantitative complementarity of wave-particle duality},
  journal = {Sci. Adv.},
  volume  = {7},
  pages   = {eabi9268},
  year    = {2021},
  doi     = {10.1126/sciadv.abi9268}
}

@article{Jakob2010,
  author  = {Jakob, M. and Bergou, J. A.},
  title   = {Quantitative complementarity relations in bipartite systems: Entanglement as a physical reality},
  journal = {Opt. Commun.},
  volume  = {283},
  pages   = {827--830},
  year    = {2010},
  doi     = {10.1016/j.optcom.2009.10.044}
}

@article{Qian2020,
  author  = {Qian, X.-F. and Konopka, K. and Eberly, J. H.},
  title   = {Turning off quantum duality},
  journal = {Phys. Rev. Research},
  volume  = {2},
  pages   = {012016},
  year    = {2020},
  doi     = {10.1103/PhysRevResearch.2.012016}
}

@article{Gottesman2001,
  author  = {Gottesman, Daniel and Kitaev, Alexei and Preskill, John},
  title   = {Encoding a qubit in an oscillator},
  journal = {Phys. Rev. A},
  volume  = {64},
  pages   = {012310},
  year    = {2001},
  doi     = {10.1103/PhysRevA.64.012310}
}

@article{Fabre2020,
  author  = {Fabre, N. and Maltese, G. and Appas, F. and Felicetti, S. and Ketterer, A. and Keller, A. and Coudreau, T. and Baboux, F. and Amanti, M. I. and Ducci, S. and Milman, P.},
  title   = {Generation of a time-frequency {G}ottesman-{K}itaev-{P}reskill state in an integrated biphoton frequency comb},
  journal = {Phys. Rev. A},
  volume  = {102},
  pages   = {012607},
  year    = {2020},
  doi     = {10.1103/PhysRevA.102.012607}
}

@article{Descamps2024,
  author  = {Descamps, E. and Keller, A. and Milman, P.},
  title   = {{G}ottesman-{K}itaev-{P}reskill encoding in continuous modal variables of single photons},
  journal = {Phys. Rev. Lett.},
  volume  = {132},
  pages   = {170601},
  year    = {2024},
  doi     = {10.1103/PhysRevLett.132.170601}
}

@article{Lindblad1976,
  author  = {Lindblad, G.},
  title   = {On the generators of quantum dynamical semigroups},
  journal = {Commun. Math. Phys.},
  volume  = {48},
  pages   = {119--130},
  year    = {1976},
  doi     = {10.1007/BF01608499}
}

@article{Baumgratz2014,
  author  = {Baumgratz, T. and Cramer, M. and Plenio, M. B.},
  title   = {Quantifying Coherence},
  journal = {Phys. Rev. Lett.},
  volume  = {113},
  pages   = {140401},
  year    = {2014},
  doi     = {10.1103/PhysRevLett.113.140401}
}

@misc{Bloch2025,
  author  = {Bloch, Yakov},
  title   = {The time qubit},
  year          = {2025},
  eprint        = {2511.07220},
  archivePrefix = {arXiv},
  primaryClass  = {quant-ph}
}

@misc{Scholes2026a,
  author  = {Scholes, Gregory D.},
  title   = {Mechanism of wavefunction collapse in measurements of separated quantum subsystems},
  year          = {2026},
  eprint        = {2605.20111},
  archivePrefix = {arXiv},
  primaryClass  = {quant-ph}
}

@misc{Scholes2026b,
  author  = {Scholes, Gregory D.},
  title   = {Measurement of quantum states and nonlocal correlations},
  year          = {2026},
  eprint        = {2605.21243},
  archivePrefix = {arXiv},
  primaryClass  = {quant-ph}
}

@misc{Scholes2026c,
  author  = {Scholes, Gregory D.},
  title   = {Measurements on the separated subsystems of an entangled state},
  year          = {2026},
  eprint        = {2609.04178},
  archivePrefix = {arXiv},
  primaryClass  = {quant-ph}
}

@article{Hansch2006,
  title = {{N}obel Lecture: Passion for precision},
  author = {H\"ansch, Theodor W.},
  journal = {Rev. Mod. Phys.},
  volume = {78},
  issue = {4},
  pages = {1297--1309},
  year = {2006},
  doi = {10.1103/RevModPhys.78.1297},
  publisher = {American Physical Society}
}

@article{Freedman1972,
  title = {Experimental test of local hidden-variable theories},
  author = {Freedman, Stuart J. and Clauser, John F.},
  journal = {Phys. Rev. Lett.},
  volume = {28},
  issue = {14},
  pages = {938--941},
  year = {1972},
  doi = {10.1103/PhysRevLett.28.938},
  publisher = {American Physical Society}
}

@article{Aspect1982,
  title = {Experimental realization of {E}instein-{P}odolsky-{R}osen-{B}ohm \textit{Gedankenexperiment}: A new violation of {B}ell's inequalities},
  author = {Aspect, Alain and Grangier, Philippe and Roger, G\'erard},
  journal = {Phys. Rev. Lett.},
  volume = {49},
  issue = {2},
  pages = {91--94},
  year = {1982},
  doi = {10.1103/PhysRevLett.49.91},
  publisher = {American Physical Society}
}

@article{Weihs1998,
  title = {Violation of {B}ell's inequality under strict {E}instein locality conditions},
  author = {Weihs, Gregor and Jennewein, Thomas and Simon, Christoph and Weinfurter, Harald and Zeilinger, Anton},
  journal = {Phys. Rev. Lett.},
  volume = {81},
  issue = {23},
  pages = {5039--5043},
  year = {1998},
  doi = {10.1103/PhysRevLett.81.5039},
  publisher = {American Physical Society}
}

@article{Hall2006,
  title = {{N}obel Lecture: Defining and measuring optical frequencies},
  author = {Hall, John L.},
  journal = {Rev. Mod. Phys.},
  volume = {78},
  issue = {4},
  pages = {1279--1295},
  year = {2006},
  doi = {10.1103/RevModPhys.78.1279},
  publisher = {American Physical Society}
}

@misc{Yoon2026a,
  author  = {Yoon, Tai Hyun},
  title   = {Hardware-native time-frequency {G}ottesman-{K}itaev-{P}reskill logical qubits in coherently seeded single-photon frequency combs},
  year          = {2026},
  eprint        = {2602.14461},
  archivePrefix = {arXiv},
  primaryClass  = {quant-ph}
}

@misc{Yoon2026b,
  author  = {Yoon, Tai Hyun},
  title   = {Measurement-defined control of single-particle interference},
  year          = {2026},
  eprint        = {2604.17767},
  archivePrefix = {arXiv},
  primaryClass  = {quant-ph}
}

@misc{Yoon2026c,
  author  = {Yoon, Tai Hyun},
  title   = {Single-photon time qubit: Quantum-vacuum control and complementarity of the arrow of time},
  year          = {2026},
  eprint        = {2610.27639},
  archivePrefix = {arXiv},
  primaryClass  = {quant-ph}
}

@article{Zavatta2004,
  author = {Zavatta, Alessandro and Viciani, Silvia and Bellini, Marco},
  title = {Quantum-to-Classical Transition with Single-Photon-Added Coherent States of Light},
  journal = {Science},
  volume = {306},
  number = {5696},
  pages = {660--662},
  year = {2004},
  doi = {10.1126/science.1103190},
  publisher = {American Association for the Advancement of Science}
}

@article{Einstein1935,
  author  = {Einstein, A. and Podolsky, B. and Rosen, N.},
  title   = {Can quantum-mechanical description of physical reality be considered complete?},
  journal = {Phys. Rev.},
  volume  = {47},
  pages   = {777--780},
  year    = {1935},
  doi     = {10.1103/PhysRev.47.777}
}

@book{Takesaki1970,
  author    = {Takesaki, Masamichi},
  title     = {Tomita's Theory of Modular Hilbert Algebras and its Applications},
  series    = {Lecture Notes in Mathematics},
  volume    = {128},
  publisher = {Springer-Verlag},
  address   = {Berlin, New York},
  year      = {1970},
  isbn      = {978-3-540-04917-3}
}

@article{Bisognano1975,
  author  = {Bisognano, J. J. and Wichmann, E. H.},
  title   = {On the duality condition for a {H}ermitian scalar field},
  journal = {J. Math. Phys.},
  volume  = {16},
  pages   = {985--1007},
  year    = {1975},
  doi     = {10.1063/1.522605}
}

@article{Ou1990,
  author  = {Ou, Z. Y. and Zou, X. Y. and Wang, L. J. and Mandel, L.},
  title   = {Observation of nonclassical effects in the interference of photons from two independent sources},
  journal = {Phys. Rev. Lett.},
  volume  = {65},
  pages   = {321--324},
  year    = {1990},
  doi     = {10.1103/PhysRevLett.65.321}
}

@article{Heuer2015,
  author  = {Heuer, Axel and Menzel, Ralf and Perez-Leija, Armando},
  title   = {Commutative and noncommutative quantum erasure with induced coherence},
  journal = {Phys. Rev. A},
  volume  = {92},
  pages   = {023832},
  year    = {2015},
  doi     = {10.1103/PhysRevA.92.023832}
}

@article{Englert1996,
  author  = {Englert, Berthold-Georg},
  title   = {Fringe Visibility and Which-Way Information: An Inequality},
  journal = {Phys. Rev. Lett.},
  volume  = {77},
  pages   = {2154--2157},
  year    = {1996},
  doi     = {10.1103/PhysRevLett.77.2154}
}

@book{Carmichael1993,
  author    = {Carmichael, Howard J.},
  title     = {An Open Systems Approach to Quantum Optics},
  publisher = {Springer-Verlag},
  address   = {Berlin},
  year      = {1993},
  doi     = {10.1007/978-3-540-47628-3}
}

@article{Fluhmann2019,
  author  = {Fl{\"u}hmann, Christa and Nguyen, Ting Rei and Marinelli, Matteo and Nehra, Vishal and Chiaverini, John and Home, Jonathan P.},
  title   = {Encoding a qubit in a trapped-ion mechanical oscillator with continuous-variable error correction},
  journal = {Nature},
  volume  = {566},
  pages   = {513--517},
  year    = {2019},
  doi     = {10.1038/s41586-019-0960-6}
}

@misc{Yoon2026d,
  author        = {Yoon, Tai Hyun},
  title         = {Resolving nonlocal state reduction via induced coherence in frequency-comb biphoton sources},
  year          = {2026},
  eprint        = {2610.27602},
  archivePrefix = {arXiv},
  primaryClass  = {quant-ph}
}

@book{Peskin1995,
  author    = {Peskin, Michael E. and Schroeder, Daniel V.},
  title     = {An Introduction to Quantum Field Theory},
  publisher = {Addison-Wesley},
  address   = {Reading, MA},
  year      = {1995}
}

@article{VillasBoas2025,
  title   = {Bright and Dark States of Light: The Quantum Origin of Classical Interference},
  author  = {Villas-Boas, Celso J. and M{\'a}ximo, Carlos E. and Paulino, Paulo J. and Bachelard, Romain P. and Rempe, Gerhard},
  journal = {Phys. Rev. Lett.},
  volume  = {134},
  pages   = {133603},
  year    = {2025},
  doi     = {10.1103/PhysRevLett.134.133603}
}

@article{Maccone2020,
  title   = {Quantum Measurements of Time},
  author  = {Maccone, Lorenzo and Sacha, Krzysztof},
  journal = {Phys. Rev. Lett.},
  volume  = {124},
  pages   = {110402},
  year    = {2020},
  doi     = {10.1103/PhysRevLett.124.110402}
}

\end{document}